%% file: young_mustang.tex
\documentclass[iop,numberedappendix,apj]{emulateapj}
\pdfoutput=1
\usepackage{color}
\usepackage{amssymb}
\usepackage{natbib}
\usepackage{graphicx}
\usepackage{url}
\usepackage{xcolor}
\usepackage{hyperref}
\hypersetup{colorlinks,linkcolor={blue!50!black},citecolor={blue!50!black},urlcolor={blue!80!black}}
\usepackage[tbtags]{amsmath}
\usepackage{gensymb}
\newcommand{\apgt} {\ {\raise-.5ex\hbox{$\buildrel>\over\sim$}}\ }
\newcommand{\aplt} {\ {\raise-.5ex\hbox{$\buildrel<\over\sim$}}\ }
\newcommand{\dmod}{\overrightarrow{d}_{mod}}
\newcommand{\dvec}{\overrightarrow{d}}
\newcommand{\avec}{\overrightarrow{a}}
\newcommand{\asec}{$^{\prime \prime}$}
\newcommand{\amin}{$^{\prime}$}
\newcommand{\sigT}{\mbox{$\sigma_{\mbox{\tiny T}}$}}
\newcommand{\sigV}{\mbox{$\sigma_{\mbox{\tiny V}}$}}
\newcommand{\Tcmb}{\mbox{$T_{\mbox{\tiny CMB}}$}}
\newcommand{\kB}{\mbox{$k_{\mbox{\tiny B}}$}}
\newcommand{\kBT}{\mbox{$k_{\mbox{\tiny B}}T_{\mbox{\tiny e}}$}}

\newcommand{\Lamee}{\mbox{$\Lambda_{\mbox{\scriptsize ee}}$}}

\newcommand{\Ysph}{\mbox{$Y_{\mbox{\scriptsize sph}}$}}
\newcommand{\Ysz}{\mbox{$Y_{\mbox{\tiny SZ}}$}}

\newcommand{\DA}{\mbox{$D_{\mbox{\tiny A}}$}}
\newcommand{\Pe}{\mbox{$P_{\mbox{\scriptsize e}}$}}
\newcommand{\msun}{$\rm M_{\odot}$}

\newcommand{\etal}{{\it et al.}}

\newcommand{\uJy}{\,{\rm \mu Jy} }

\newcommand{\chandra}{{\it Chandra}}

\newcommand{\xmm}{{XMM-{\it Newton}}}

\newcommand{\rxj}{RX J1347.5-1145}
\newcommand{\clj}{CL J1226.9+3332}
\newcommand{\macsa}{MACS J0647.7+7015}
\newcommand{\macsb}{MACS J1206.2-0847}
\newcommand{\macsc}{MACS J0717.5+3745}
\newcommand{\macsg}{MACS J0744.9+3927}
\newcommand{\Lx}{\mbox{$L_{\mbox{\tiny X}}$}}
\newcommand{\Tx}{\mbox{$T_{\mbox{\tiny X}}$}}
\newcommand{\te}{\mbox{$T_{\mbox{\tiny e}}$}}
\newcommand{\mec}{\mbox{$m_{\mbox{\tiny e}} c^2$}}
\newcommand{\dene}{\mbox{$n_{\mbox{\tiny e}}$}}

\newcommand{\sx}{\mbox{$S_{\mbox{\tiny X}}$}}
\newcommand{\Isz}{\mbox{$I_{\mbox{\tiny SZE}}$}}

\newcommand{\chisq}{\mbox{$\chi^{2}$}}
\newcommand{\chired}{\mbox{$\chi^{2}_{red}$}}
\newcommand{\clash}{Cluster Lensing And Supernova survey with Hubble}

\newcommand{\romb}{Romero et al.\ (in prep.)}

\newcommand{\macsbflux}{$61 \pm 21$}
\newcommand{\avg}[1]{\langle {#1} \rangle}

\slugcomment{}
\shortauthors{Young \etal}
\shorttitle{MUSTANG SZE imaging of two galaxy clusters}
\begin{document}
\title{Measurements of the Sunyaev-Zel'dovich Effect in MACS~J0647.7+7015 
and MACS~J1206.2-0847 at High Angular Resolution with MUSTANG}
\author{
  Alexander H.\ Young\altaffilmark{1,2},
  Tony Mroczkowski\altaffilmark{3},
  Charles Romero\altaffilmark{4,5},
  Jack Sayers\altaffilmark{6},
  Italo Balestra\altaffilmark{7},
  Tracy E.\ Clarke\altaffilmark{8},
  Nicole Czakon\altaffilmark{9},
  Mark Devlin\altaffilmark{1},
  Simon R.\ Dicker\altaffilmark{1},
  Chiara Ferrari\altaffilmark{10},
  Marisa Girardi\altaffilmark{7,11},
  Sunil Golwala\altaffilmark{6},
  Huib Intema\altaffilmark{12},
  Phillip M.\ Korngut\altaffilmark{6},
  Brian S.\ Mason\altaffilmark{4},
  Amata Mercurio\altaffilmark{13},
  Mario Nonino\altaffilmark{7},
  Erik D.\ Reese\altaffilmark{14},
  Piero Rosati\altaffilmark{15},
  Craig Sarazin\altaffilmark{5},
  and
  Keiichi Umetsu\altaffilmark{9}
} 
\altaffiltext{1}{Department of Physics and Astronomy, University of
  Pennsylvania, 209 South 33rd Street, Philadelphia, PA 19104, USA;
  \email{alyoung@sas.upenn.edu}}

\altaffiltext{2}{NASA Postdoctoral Fellow at NASA Goddard Space Flight
  Center, 8800 Greenbelt Rd, Greenbelt, MD 20771, USA}

\altaffiltext{3}{National Research Council Fellow at the U.S.\ Naval
  Research Laboratory, 4555 Overlook Ave SW, Washington, DC 20375,
  USA}

\altaffiltext{4}{National Radio Astronomy Observatory, 520 Edgemont
  Rd.  Charlottesville VA 22903, USA} 

\altaffiltext{5}{Department of Astronomy, University of Virginia,
  P.O. Box 400325, Charlottesville, VA 22901, USA}

\altaffiltext{6}{Department of Physics, Math, and Astronomy,
  California Institute of Technology, 1200 East California Blvd,
  Pasadena, CA 91125}

\altaffiltext{7}{INAF - Osservatorio Astronomico di Trieste, via
  G. B. Tiepolo 11, I-34143 Trieste, Italy}

\altaffiltext{8}{U.S.\ Naval Research Laboratory, 4555 Overlook Ave
  SW, Washington, DC 20375, USA}

\altaffiltext{9}{Institute of Astronomy and Astrophysics, Academia
  Sinica, P.O. Box 23-141, Taipei 10617, Taiwan}

\altaffiltext{10}{Laboratoire Lagrange, UMR 7293, Universit\'{e} de
  Nice Sophia-Antipolis, CNRS, Observatoire de la C\^{o}te d'Azur,
  06300 Nice, France}

\altaffiltext{11}{Dipartimento di Fisica Dell'Universit\'{a} degli
  Studi di Trieste - Sezione di Astronomia, via Tiepolo 11, I-34143
  Trieste, Italy}

\altaffiltext{12}{National Radio Astronomy Observatory, P.O. Box O,
  1003 Lopezville Road, Socorro, NM 87801-0387, USA}

\altaffiltext{13}{INAF - Osservatorio Astronomico di Capodimonte, via
  Moiariello 16, I-80131 Napoli, Italy}

\altaffiltext{14}{Department of Physics, Astronomy, and Engineering, 
  Moorpark College, 7075 Campus Rd., Moorpark, CA 93021, USA} 

\altaffiltext{15}{Dipartimento di Fisica e
  Scienze della Terra, Universit\'a di Ferrara, Via Saragat, 1,
  I-44122, Ferrara, Italy} 

\begin{abstract}
We present high resolution (9$^{\prime \prime}$) imaging of the
Sunyaev-Zel'dovich Effect (SZE) toward two massive galaxy clusters,
MACS~J0647.7+7015 ($z=0.591$) and MACS~J1206.2-0847 ($z=0.439$). We
compare these 90~GHz measurements, taken with the MUSTANG receiver on
the Green Bank Telescope, with generalized Navarro-Frenk-White (gNFW)
models derived from Bolocam 140~GHz SZE data as well as maps of the
thermal gas derived from {\it Chandra} X-ray observations.  For MACS
J0647.7+7015, we find a gNFW profile with core slope parameter
$\gamma= 0.9$ fits the MUSTANG image with
\mbox{$\chi^{2}_{red}=1.005$} and probability to exceed (PTE) =
0.34. For MACS~J1206.2-0847, we find $\gamma=0.7$,
\mbox{$\chi^{2}_{red}=0.993$}, and PTE = 0.70. In addition, we find a
significant ($>$3-$\sigma$) residual SZE feature in MACS~J1206.2-0847
coincident with a group of galaxies identified in VLT data and
filamentary structure found in a weak-lensing mass reconstruction. We
suggest the detected sub-structure may be the SZE decrement from a low
mass foreground group or an infalling group. GMRT measurements at
610~MHz reveal diffuse extended radio emission to the west, which we
posit is either an AGN-driven radio lobe, a bubble expanding away from
disturbed gas associated with the SZE signal, or a bubble detached and
perhaps re-accelerated by sloshing within the cluster. Using the
spectroscopic redshifts available, we find evidence for a foreground
($z=0.423$) or infalling group, coincident with the residual SZE
feature.
 
\end{abstract}

\section{INTRODUCTION}
\label{sec:intro}

Clusters of galaxies are the largest gravitationally bound systems in
the Universe and encompass volumes great enough to be considered
representative samples of the Universe at large. By mass, clusters comprise
dark matter ($\sim$85$\%$), hot plasma known as the intra-cluster medium
(ICM; $\sim$12$\%$), and a few percent stars and galaxies
\citep{sarazin2002}.

The diverse matter content of clusters provides a wide range of
observables across the electromagnetic spectrum. X-ray and
millimeter-wave Sunyaev-Zel'dovich Effect (SZE) observations measure
the thermodynamic properties of the ICM, such as density and
temperature, which provides insight into the formation history and
evolution of the cluster as well as its current dynamical state. Radio
observations have discovered diffuse synchrotron emission in many
galaxy clusters, typically associated with merger-induced shock
fronts, turbulence, or Active Galactic Nucleus (AGN) activity (e.g,
\citealt{vanweeren2011,cassano2012}). Optical observations reveal the
individual galaxy population and, through dynamical and lensing
studies, allow us to infer the cluster mass distribution.

\input{tab1.tex}

The SZE is a distortion of the Cosmic Microwave Background (CMB)
caused by inverse Compton scattering of photons off the electrons of
the hot ICM trapped in the gravitational potential well of
clusters. The SZE is directly proportional to the electron pressure of
the ICM integrated along the line of sight \citep{Sunyaev1972}.
Measurements of the SZE on small spatial scales in galaxy clusters
provide a powerful probe of astrophysical phenomena
\citep[e.g.,][]{kitayama2004, korngut2011}. For reviews of the SZE and
its scientific applications, see \citet{birkinshaw1999} and
\citet{carlstrom2002}.

In this work, we present high-resolution SZE measurements of two
galaxy clusters, \macsa\ and \macsb, taken with the Multiplexed
Squid/TES Array at Ninety Gigahertz (MUSTANG). We carry out a
multi-wavelength investigation using the comprehensive data sets
provided by the \clash\ (CLASH) program \citep{postman2012}. The X-ray
measured properties of \macsa\ (MACSJ0647.7) and \macsb\ (MACSJ1206.8)
from \citet{mantz2010b} are summarized in Table~\ref{tbl:clusprops}.

The organization of this paper is as follows. In \S\ref{sec:obs} we
discuss the MUSTANG, X-ray, and Bolocam observations and data
reduction. In \S\ref{sec:analysis} and \S\ref{sec:models}, we discuss
the ICM modeling and least-squares fitting procedure used in the
combined analysis of the MUSTANG and Bolocam data. The results are
discussed and summarized in \S\ref{sec:results}. Throughout this
paper, we adopt a flat, $\Lambda$-dominated cosmology with
$\Omega_{\mbox{\tiny \rm M}} = 0.3$, $\Omega_\Lambda = 0.7$, and $H_0
= 70$ km s$^{-1}$ Mpc$^{-1}$ consistent with {\it Planck} results
\citep{planck2013xvi}. At the redshifts of \macsa\ ($z=0.591$) and
\macsb\ ($z=0.439$), 1\asec\ corresponds to 6.64~kpc and 5.68~kpc,
respectively.

\section{OBSERVATIONS AND DATA REDUCTION}
\label{sec:obs}

\subsection{The CLASH Sample}
In this paper, we present MUSTANG observations of \macsa\ and
\macsb. Basic characteristics of these clusters are summarized in
Tables~\ref{tbl:clusprops} \& \ref{tbl:obs}, as part of an ongoing
program to provide high-resolution SZE images of the CLASH clusters
accessible from MUSTANG's location on the Green Bank Telescope (GBT;
\citealt{jewell2004}). The 25 clusters in CLASH have comprehensive
multi-wavelength coverage, including deep 16-band HST optical imaging,
relatively low resolution SZE measurements, and X-ray observations
with \chandra\ and \xmm. These clusters are generally dynamically
relaxed, span redshifts from $0.2 \aplt z \aplt 0.9$, and masses from
$3\times10^{14}\aplt M_{500}/$\msun$\aplt 2\times10^{15}$. For a
comprehensive description of the CLASH sample and selection criteria
see \citet{postman2012}.

\input{tab2.tex}

\subsection{MUSTANG}

MUSTANG is a 64-pixel array of Transition Edge Sensor (TES) bolometers
spaced at $0.6f\lambda$ operating at 90~GHz on the 100-meter
GBT. MUSTANG has an instantaneous field of view (FOV) of 42\asec\ and
angular resolution of 9\asec.  For more information about MUSTANG,
refer to \citet{dicker2008}.

MUSTANG has measured the SZE at high resolution in several galaxy
clusters to date, including \rxj, \clj, \macsc, \macsg, and
Abell~1835.  MUSTANG observations confirmed ($>$13-$\sigma$) the
presence of merger activity in
\rxj\ \citep{mason2010,korngut2011,ferrari2011} that was hinted at by
observations with the Nobeyama 45~m telescope \citep{komatsu2001} and
the 30~m IRAM telescope \citep{pointecouteau1999}.
\citet{korngut2011} used MUSTANG data to discover a shock in
\macsg\ that was previously undetected.  In \macsc,
\citet{mroczkowski2012} used MUSTANG data to report a pressure
enhancement due to shock-heated gas immediately adjacent to extended
radio emission.

The MUSTANG observations and data reduction in this work largely
follow the procedure described in \citet{mason2010} and
\citet{korngut2011}. We direct the telescope in a Lissajous daisy scan
pattern with seven pointing centers surrounding the cluster core.
This mosaic provides deep, uniform coverage in the cluster core and
falls off steeply beyond a radius of $\sim$30\asec.

During observations, nearby bright compact radio sources were mapped
once every 30~minutes to track changes in the beam profile including
drifts in telescope gain and pointing offsets. Typically, if there was
a substantial ($\sim$20$\%$) drop in the peak of the beam profile, or if
the beam width exceeded 10\asec, we re-derived the GBT active surface
corrections using an out-of-focus (OOF) holography technique
\citep{nikolic2007}. We used the blazar JVAS~0721+7120 for \macsa\ and
the quasar JVAS~1229+0203 for \macsb\ to determine these gains and
focusing corrections. Planets or stable quasars including Mars,
Saturn, and 3C286 \citep{agudo2012} were mapped at least once per
observation session to provide absolute flux calibration. Fluxes for
planets were calculated based on brightness temperatures from WMAP
observations \citep{weiland2011}. The absolute flux of the data is
calibrated to an accuracy of 10\%. Throughout this work, we ignore the
systematic uncertainty from the absolute flux calibration and quote
only the statistical uncertainties.

The MUSTANG data are reduced using a custom IDL pipeline. The
bolometric timestreams are high-pass filtered by subtracting a high
order Legendre polynomial determined by the scan speed of the
telescope. For a typical 300~s scan, and 40\asec~s$^{-1}$ scan speed,
we choose a $~\sim$100$^{\rm th}$-order Legendre polynomial,
corresponding to a cutoff frequency of $\sim$0.3~Hz. In order to
remove atmospheric noise on large angular scales, we subtract the mean
measurement from all detectors for each sample in time.  This also
removes astronomical signals on angular scales larger than the FOV of
the instrument ($\approx$42\asec).

The standard deviation, $\sigma$, of each individual detector timestream
is computed, and a corresponding weight, $w$, is determined according to $w
= 1/\sigma^{2}$. To produce a ``signal map'', the timestreams are
binned into 1\asec~$\times$~1\asec\ spatial pixels and smoothed with
the MUSTANG point spread function (PSF), or beam. We compute the
weight for each pixel of the smoothed data map to produce a ``weight
map''. We multiply the signal map by the square root of the weight map
to generate a map in units of S/N - the ``SNR map''.

We generate an independent ``noise map'' by flipping the sign of
measurements from every other scan and binning the data into a grid
with the same pixel size as the signal map. As we do for the signal
map, we use the pixel weights to convert the noise map to units of
S/N, referred to as a ``noise SNR map''. We define a scale factor,
$\sigma_{N}$, as the standard deviation of the noise SNR map. For an
ideal noise distribution, $\sigma_{N}=1$.  We can therefore use
$\sigma_{N}$ as a normalization factor to account for ``non-ideal''
noise features, such as correlations between detectors. Typically, we
find $\sigma_{N}\approx1.5$, which means that the timestream-based
weight maps are under-estimating the noise.

\subsection{Bolocam}

Bolocam is a 144-pixel bolometer array at the Caltech Submillimeter
Observatory (CSO) capable of operating at 140 and 268~GHz, with
resolutions of 31\asec\ and 58\asec, respectively, and an
instantaneous FOV 8\amin\ in diameter. For more details on the Bolocam
instrument see \citet{haig2004}.

As part of a larger cluster program \citep{sayers2013,czakon2014},
Bolocam was used to obtain high significance SZE images of
\macsa\ (S/N~=~14.4) and \macsb\ (S/N~=~21.7). In this work, we make use
of these Bolocam data to constrain bulk models of the SZE emission
based on generalized Navarro-Frenk-White (gNFW) pressure profiles
\citep{nagai2007}, including the specific case of the ``universal
pressure profile''\citep[][hereafter A10]{arnaud2010}. The model
fitting procedure is described in \S\ref{sec:bolomodels}, and the
details of the Bolocam data, along with its reduction are given in
\citet[][hereafter S13]{sayers2013} and \citet{czakon2014}.

\subsection{Chandra}
\label{sec:chandra}

Archival \chandra\ X-ray data were reduced using
CIAO\footnote{http://cxc.harvard.edu/ciao/} version 4.5 with
calibration database (CALDB) version 4.5.5. \macsa\ was observed for a
total exposure time of 39~ks (ObsIDs 3196 and 3584).  \macsb\ was
observed for 24~ks (ObsID 3277). For details on the X-ray data
processing see \citet{reese2010}.

\section{ICM ANALYSIS}
\label{sec:analysis}

The thermal SZE intensity is described by 
\begin{equation}
\label{eq:tsze}
\frac{\Delta \Isz}{I_{0}}= g(\nu, \te)y,
\end{equation}
where $\nu$ is the observed frequency, $T_{e}$ is the electron
temperature, $y$ is the Compton-$y$ parameter (described below), and
the primary CMB surface brightness is $I_{0} = 2(\kB
\Tcmb)^{3}(hc)^{-2} = 2.7033\times10^{8}$~Jy~sr$^{-1}$. The function
$g(\nu,\te)$ describes the frequency dependence of the thermal SZE
\citep{carlstrom2002} and includes the relativistic corrections of
\citet{itoh1998} and \citet{itoh2004}.  At 90~GHz, the SZE manifests
as a decrement in the CMB intensity.

The frequency-independent Compton-$y$ parameter is defined as
\begin{equation}
  \label{eq:compy}
  y \equiv \frac{\sigT}{\mec} \int \! \dene \kB \te \,d\ell,
\end{equation}
where $\sigT$ is the Thomson cross section, \mec\ is the electron rest
energy, and the integration is along the line of sight
$\ell$. Therefore, by the ideal gas law, the SZE intensity is
proportional to the ICM electron pressure $\Pe=\dene \kB \te$
integrated along the line of sight. The total SZE signal, integrated
within an aperture $\theta=R/D_{A}$, is often expressed in units of
solid angle, where \Ysz[sr] $ = \int \! y \, d\Omega$ or in distance units where
$Y [$Mpc$^{2}]=\Ysz D_{A}^{2}$.

The X-ray surface brightness (in units of counts
cm$^{-2}$~s$^{-1}$~sr$^{-1}$) is
\begin{equation*} \label{eq:sx}
  \sx = \frac{1}{4\pi(1+z)^{3}}\int \! \dene^{2}\Lamee(\te,Z)d\ell
\end{equation*}
where \Lamee$(\te,Z)$ is the X-ray cooling function, and $Z$ is the
abundance of heavy elements relative to that in the Sun. Assuming the
temperature is constant along the line of sight, 
\begin{equation} \label{eq:nx}
  \dene \approx \sqrt{\frac{4\pi(1+z)^{3}\sx}{\Lamee(\te,Z)\ell}}.
\end{equation}

We approximate Equation~\ref{eq:compy} as
$y\approx\sigT/(\mec)\dene\kB\te\ell$ and use Equation~\ref{eq:nx} to
derive from the X-ray data a ``pseudo''-$y$ value\footnote{The
  ``pseudo'' distinction is used because $\ell$ is not constrained by
  the X-ray data alone.}, given by

\begin{equation}
  y= \frac{\sigT \kB \te}{\mec} 
  \sqrt{\frac{4\pi (1+z)^{3}\sx \ell}{\Lamee(\te,Z)}}.
  \label{eq:ixray} 
\end{equation}

We use a measurement of the integrated Compton-$y$ ($\Ysz D_{A}^{2}$)
within $R<1$\amin\ from Bolocam to infer $\ell$ and normalize the
X-ray pseudo-$y$ map accordingly. This assumes that $\ell$ is constant
radially, which is a reasonable approximation for typical cluster
density profiles \citep[see][]{mroczkowski2012}. Additionally, we
assume both clusters have an isothermal temperature distribution with
the \kBT\ values determined by X-ray spectroscopy.  We note that for
each of these clusters, the assumption of an isothermal distribution
within $r\aplt 120$\asec\ is reasonable based on the relatively flat
radial temperature profiles given in the Archive of Chandra Cluster
Entropy Profile Tables (ACCEPT) database \citep{cavagnolo2009}.

\input{fig1.tex}

Several measurements have shown that the pressure of the ICM is well
described by a gNFW pressure profile (e.g., \citealt{mroczkowski2009}; A10;
\citealt{plagge2010}; \citealt{planck2013v}; S13).  In this model, the
pressure (in units of $P_{500}$) is
\begin{equation}
  \label{eq:gNFW}
  \tilde{P}(X) =
  \frac{P_{0}}{(C_{500}X)^{\gamma}[1+(C_{500}X)^{\alpha}]^{(\beta-\gamma)/\alpha}},
\end{equation}
where $X=R/R_{500}$\footnote{$R_{\Delta}$ is defined as the radius at
  which the mean interior mass density of a cluster is $\Delta$ times
  the critical density of the Universe at the redshift of the cluster:
  $M_{\Delta}=(4\pi/3)\Delta\rho_{c}R_{\Delta}^{3}$.}, $C_{500}$ is
the concentration parameter, often given in terms of the scale radius
$R_{s}$ ($C_{500}=R_{500}/R_{s}$), $P_{0}$ is the normalization
factor, $\gamma$ is the inner slope ($r << R_{s}$), $\alpha$ is the
intermediate slope ($r\sim R_{s}$), and $\beta$ is the outer slope ($r
>> R_{s}$). $P_{500}$ is defined in Equation~\ref{eq:P500}.

\input{tab3.tex}

In this work, we focus on the gNFW fit results from A10 and S13. The
gNFW model parameters for the respective ensemble samples, in addition
to subsets defined according to cluster morphology, are given in
Table~\ref{tbl:nfwmods}.  We also include the best-fit parameters for
\macsa\ and \macsb\ determined in \S\ref{sec:results}. Pressure
profiles for each of these models, scaled based on the values of
$P_{500}$, $R_{500}$, and $z$ given in Table~\ref{tbl:clusprops} for
each cluster, are shown in Figure~\ref{fig:profiles}. We also include
plots of the spherically integrated Compton-$y$, \Ysph$(<R)$, given by
\begin{equation*}\label{eq:ysph}
Y_{\rm sph}(<R)=\frac{4\pi\sigT}{\mec}\int_0^R P(r)r^{2}dr.
\end{equation*}
As in A10, we express \Ysph\ in units of $Y_{500}$, where
\begin{equation}\label{eq:y500}
  Y_{500}=\frac{\sigT}{\mec}\frac{4\pi}{3}R_{500}^{3}P_{500}, 
\end{equation}
and 
\begin{equation}\label{eq:P500}
P_{500}=\left(3.68 \times 10^{-3} \frac{\rm keV}{\rm cm^{3}}\right)
  \left(\frac{M_{500}}{10^{15}M_{\odot}}\right)^{2/3} E(z)^{8/3},
\end{equation}
where $E^{2}(z)=\Omega_{M}(1+z)^3+\Omega_{\Lambda}$ (see
\citealt{nagai2007}; A10; S13).  The values of $Y_{500}$ and $P_{500}$
from Equations~\ref{eq:y500} \& \ref{eq:P500}, respectively,
are derived from the cluster properties reported in
\citealt{mantz2010b} and summarized in Table~\ref{tbl:clusprops}.

\section{MODEL FITTING}
\label{sec:models}

While MUSTANG provides high-resolution imaging, the angular transfer
function falls off steeply beyond the instrument FOV
($\approx$42\asec\ $= 255$~kpc at $z=0.5$).  Bolocam has a lower
resolution, but a larger FOV and therefore is sensitive to the bulk
SZE signal on larger angular scales (beyond $\sim$10\amin). A combined
Bolocam+MUSTANG model-fitting approach allows us to place better
constraints on the ICM characteristics over the full range of angular
scales probed by both instruments. In this work, we present the first
steps toward a robust joint-fitting procedure.

\subsection{Fitting Algorithm}
\label{sec:fitting}
We begin by constructing a model map in units of Jy~beam$^{-1}$
smoothed to MUSTANG resolution. We simulate an observation of the
model by injecting noise from real observations and then processing
the mock observation through the MUSTANG mapmaking pipeline. By
subtracting the injected noise from the output map we obtain a
filtered model map without residual noise. Examples of these
post-processed model maps are presented in \S\ref{sec:results}.

To fit the filtered model maps to the data in the map domain we use
the general linear least squares fitting approach from \emph{Numerical
  Recipes}~\citep{press1992}, outlined briefly below.

We construct an $N\times M$ design matrix $\mathbf{A}$, where each
element $A_{ij}$ corresponds to a model component (e.g., a point
source or gNFW model) $X_{j}$ evaluated at map pixel $x_{i}$.  In this
work, we allow a single free parameter for each model component, a
scalar amplitude, $a_{j}$.  We call the M-length vector of amplitudes
$\avec$ and define a model vector,
\begin{equation*}
  \dmod = \mathbf{A}\avec.
\end{equation*}
The goodness of fit statistic, \chisq, is given by
\begin{equation*}
\label{eqn:linearchi}
\chi^2 = (\overrightarrow{d} - \overrightarrow{d}_{mod})^T
\mathbf{N}^{-1} (\overrightarrow{d} - \overrightarrow{d}_{mod}),
\end{equation*}
where $\dvec$ represents the measured values of each map pixel and
$\mathbf{N}$ is the noise covariance matrix, where
\begin{equation*}
  N_{ij} = <n_i n_j> - <n_i> <n_j>.
\end{equation*}
Here, $\overrightarrow{n}$ is taken to be pixel values of a noise map,
and the covariance matrix is calculated using the ensemble average
over statistically identical noise realizations. Given that our
detector noise is dominated by phonon noise, pixel noise is largely
uncorrelated, so we therefore take the noise covariance matrix
$\mathbf{N}$ to be diagonal. Residual atmospheric noise coupled with
slight correlations between detectors will contribute off-diagonal
elements to $\mathbf{N}$. These terms are on average $3\%$ of the
magnitude of the diagonal terms and ignored in this procedure for
computational simplicity. The best-fit amplitudes, corresponding to the
minimum \chisq, are then
\begin{equation*}
\label{eqn:modelamps}
  \avec = (\mathbf{A}^T \mathbf{N}^{-1} \mathbf{A})^{-1} \mathbf{A}^T
  \mathbf{N}^{-1} \overrightarrow{d}.
\end{equation*}
The parameter uncertainties $\sigma^{2}(a_{k})$ are given by the
diagonal elements of the parameter covariance matrix
$(\mathbf{A}^{T} \mathbf{N}^{-1} \mathbf{A})^{-1}$.

We perform the fits over a region within 1\amin\ of the cluster
centers. This scale is chosen to match the MUSTANG angular transfer
function and we find that the results do not change significantly for
fits using larger regions. Given the 1\asec~$\times$~1\asec\ map
pixels, this yields roughly $\pi(60)^{2} = 11,310$ degrees of freedom,
minus the number of model components we include in each fit. The
probability to exceed $\chi^2$ (PTE) reflects the probability
that deviations between the given model and the data, at least as large
as those observed, would be seen by chance, assuming the model is
correct.

\subsection{Bolo+MUSTANG gNFW profiles}
\label{sec:bolomodels}

The Bolocam gNFW profiles are derived following the fitting procedure
in \citet{sayers2011} and \citet{czakon2014}, which we summarize
briefly below.

First, the gNFW profile is used to obtain a 3-dimensional model of the
SZE. Next, this model is projected to 2-dimensions, scaled in angular
size according to the cluster redshift, and convolved with both the
Bolocam PSF and the transfer function of the Bolocam data
processing. The result is then compared to the Bolocam image in order
to obtain the best-fit parameters of the gNFW profile. For these fits,
we use Equation~\ref{eq:tsze} to convert the Bolocam brightness images
to units of Compton-$y$. We include the relativistic corrections of
\citet{itoh1998} and \citet{itoh2004}, assuming the isothermal
temperature given in Table~\ref{tbl:clusprops}.

\input{fig2.tex}

Following the above procedure, we fit the Bolocam data with gNFW
profiles spanning a range of fixed $\gamma$ values from 0 to 1.5. For
generality, we fit elliptical models to the Bolocam data, although we
note that these models produce axial ratios that are close to 1 (i.e., the
elliptical models are nearly circular). For each profile, we assume
the A10 ``universal profile'' values $\alpha=1.05$ and
$\beta=5.49$. The normalization $P_{0}$, centroid, and scale radius
$R_{s} = R_{500}/C_{500}$ are allowed to float. These best-fit
pressure profiles to Bolocam are shown in
Figure~\ref{fig:bolomodels}. The integrated Compton-$y$ profiles are
also shown.

We compare each of these models to the MUSTANG data as described in
\S\ref{sec:fitting}. We choose a grid over $\gamma$ values because
$\gamma$ defines the inner slope of the ICM profile where we expect
MUSTANG to be most sensitive. From the grid of best-fit models to the
Bolocam data, the model with the best fit to the MUSTANG data is
selected as the overall best fit, referred to as the Bolo+MUSTANG
model. Effectively, the Bolocam data constrain the values of $P_{0}$
and $C_{500}$ (for fixed $\gamma$, $\alpha$, and $\beta$), and the MUSTANG
data constrain the value of $\gamma$. 

\section{RESULTS}
\label{sec:results}

\subsection{MACS~J0647.7+7015}

\input{fig3.tex}

\macsa, discovered during the Massive Cluster Survey (MACS;
\citealt{ebeling2001}), is a seemingly relaxed massive system at
$z=0.591$, but contains multiple cD galaxies \citep[see][]{hung2012},
which may indicate ongoing merger activity \citep{mann2012}.
Figure~\ref{fig:m0647composite} shows a composite image of
\macsa\ including optical, strong-lensing, and X-ray images. The mass
distribution from the strong-lensing analysis \citep{zitrin2011} is
doubly peaked and elongated in the E-W direction.  The X-ray emission
measured by \chandra\ shows similar elongation as does the SZE flux
measured by MUSTANG.

\input{fig4.tex}

The MUSTANG map of \macsa\ is shown in Figure~\ref{fig:m0647snr}.  The
peak SZE flux is $-121 \pm 16 \uJy$~beam$^{-1}$.  The measured decrement
($>$3-$\sigma$) encompasses an elongated region approximately
25\asec~$\times$~38\asec.  The total SZE flux measured by MUSTANG,
within the region with $>$3-$\sigma$ significance of the decrement, is
$-535 \pm 38 \uJy$. 

\input{fig5.tex}

Figure~\ref{fig:m0647xray} shows the pseudo-$y$ template derived from
X-ray measurements according to Equation~\ref{eq:ixray}. Normalizing
the integrated pseudo Compton-$y$ based on the Bolocam flux as
described in \S\ref{sec:analysis} yields an effective depth
$\ell=1.4$~Mpc.

Following the procedure outlined in \S\ref{sec:analysis}, we
determine the thermal SZE model that best simultaneously describes the
MUSTANG and Bolocam data to be a gNFW profile with 
\begin{equation}
[P_{0}, C_{500}, \gamma, \alpha, \beta] = [0.54, 0.29, 0.90, 1.05, 5.49],
\end{equation}
with a ratio between major and minor axes of 1.27 and position angle
$-$174$\degree$~E~of~N, hereafter referred to as the $\gamma=0.9$, or
G9, model. $C_{500}$ is computed from the geometric mean of the major
and minor axes and using $R_{500}$ from
Table~\ref{tbl:clusprops}. Figure~\ref{fig:m0647gamma} shows the
calculated reduced $\chi^{2}$ ($\chi^{2}_{red}=\chisq/$DOF) and PTE
as a function of the fixed $\gamma$ value. The G9 model gives
$\chisq$/DOF~$ = 11378/11314$ with PTE = 0.34 (see
Table~\ref{tbl:fitresults}).

\input{fig6.tex}

The X-ray pseudo-SZE and G9 model for \macsa, after being filtered
through the MUSTANG pipeline, are shown in
Figure~\ref{fig:m0647inmod}.  Also shown are the azimuthally averaged
radial profiles. The X-ray flux is concentrated on smaller scales and
passes through the MUSTANG pipeline with less attenuation compared to
the gNFW models, which have shallower profiles extending to larger
radii. The filtered G9 flux peak is offset slightly north of the X-ray
peak. The radially averaged profiles from the filtered maps are fairly
consistent between all three data sets.

\input{fig7.tex}

\subsubsection{Discussion}\label{sec:m0647dis}

In \macsa, we find good agreement between the MUSTANG high-resolution
SZE image and the X-ray and Bolocam measurements. We summarize the
results from the fitting procedure in Table~\ref{tbl:fitresults}. The
$\gamma=0.9$ gNFW model best fits the MUSTANG data with a PTE of 0.34,
whereas the A10 and pseudo-SZE are less favored (PTE~$\leq0.23$).

The compact positive sources in Figure~\ref{fig:m0647snr} are
significant ($>$3-$\sigma$) even after accounting for the lower
observing coverage outside the cluster core, however, we find no
counterparts for these sources in any other data set. In computing the
significances we have assumed that the MUSTANG map-domain noise
follows a Gaussian distribution within a 2\amin~radius, which we
verified by inspecting the histogram of the noise map for \macsa. High
resolution radio observations were not obtained for \macsa\ so
spectral coverage close to 90~GHz is limited. We take jackknives of
the data, split into four equal integration times, and the sources
appear with similar flux in each segment, which is unlikely for an
artifact. Therefore, it is possible that these are yet unidentified
objects such as lensed high-$z$ dusty galaxies or shallow spectrum
AGNs, which may be confirmed by future observations with high
resolution coverage near 90~GHz.

\input{tab4.tex}
\subsection{MACS~J1206.2-0847}\label{sec:m1206}

\input{fig8.tex}

\macsb\ is a mostly relaxed system at $z=0.439$ that has been studied
extensively through X-ray, SZE, and optical observations (e.g.,
\citealt{ebeling2001, ebeling2009, gilmour2009, umetsu2012,
  zitrin2012, biviano2013}; S13). A composite image with the
multi-wavelength data is shown in Figure~\ref{fig:m1206composite}. We
include high resolution 610~MHz data from the Giant Metrewave Radio
Telescope (GMRT; project code 21\_017). These data reveal extended
radio emission to the west of the $\sim$0.5~Jy central AGN.

\input{fig9.tex}

The MUSTANG SZE map of \macsb\ is shown in
Figure~\ref{fig:m1206snr}. The majority of the SZE decrement extends
to the northeast and is contaminated by emission from the
central AGN.

\input{fig10.tex}

The X-ray pseudo-$y$ and SZE decrement measured by Bolocam are shown
in Figure~\ref{fig:m1206xray}.  The Bolocam normalization of the
pseudo-$y$ map yields an effective depth $\ell = 2.0$ Mpc.

The BCG ($\alpha_{\rm J2000} =12^{\rm h}06^{\rm m}12.1^{\rm
  s},$~$\delta_{\rm J2000}=-08\degree48$\amin$3$\asec) in
\macsb\ harbors a radio-loud AGN that is detected by MUSTANG at high
significance (S/N~$>4$). Using a spatial template derived from the
MUSTANG map, we construct a compact source model and allow the
amplitude to float in the joint fits with bulk SZE models, in order to
account for the degeneracy between the co-spatial positive emission
and SZE decrement. AGN brightness is generally represented as a power
law with frequency, given by
\begin{equation}
\log(S(\nu)[{\rm mJy}]=\alpha \log(\nu[{\rm MHz}]))+\beta 
\end{equation}
where $\alpha$ is the spectral index and $\beta$ is the
abscissa. Extrapolating from low frequency ($\nu < 1.4$~GHz)
measurements, SPECFIND V2.0 \citep{vollmer2010} predicts $\alpha=
-1.26\pm 0.1$ and $\beta = 6.2\pm 0.2$, or a 90~GHz flux of
$S_{90}=879\pm253 \mu$Jy. By way of comparison, our joint fit results give
$S_{90}=584-765~\mu$Jy, summarized in Table~\ref{tbl:ptsrc}.

\input{tab5.tex}

\input{fig11.tex}

Figure~\ref{fig:m1206gamma} shows the goodness of fit statistics for
the gNFW + point source model fitting. With \chired~=~0.993 and
PTE~=~0.70, the best fit model is a gNFW with
\begin{equation}
[P_{0}, C_{500}, \gamma, \alpha, \beta] = [1.13, 0.41, 0.70, 1.05, 5.49],
\end{equation}
with a ratio between major and minor axes of 1.02 and position angle
$-$13$\degree$~E~of~N, hereafter G7 (see
Table~\ref{tbl:fitresults}). The filtered G7 and pseudo-SZE models are
shown in Figure~\ref{fig:m1206inmod}. The Bolocam model is much more
extended than the X-ray and is subsequently filtered the most by the
MUSTANG transfer function. The pseudo-SZE model shows a much higher
peak after filtering, but diminishes rapidly with radius.
 
\input{fig12.tex}

After subtracting the point source and G7 model, we find a
$3$-$\sigma$ residual feature in \macsb\ (see
Figure~\ref{fig:m1206snr}). The $3$-$\sigma$ contour encompasses a
73~arcsecond$^{2}$ (2~kpc$^{2}$) region with an integrated flux of
$-$\macsbflux~$\uJy$. Using Equation~\ref{eq:tsze} we calculate the
integrated Compton-$y$, $\Ysz D_{A}^{2} = 7.3 \times
10^{-7}$~Mpc$^{2}$ (see Table~\ref{tbl:m1206fluxes}).

\input{tab6.tex}

\subsubsection{Discussion}\label{sec:m1206dis}

The Bolo+MUSTANG G7 model for \macsb\ provides a good fit to the
MUSTANG data when a point source model for the central AGN is
included. However, since the point source is co-spatial with the SZE
decrement and we allow the amplitude to float, a model with a steeper
core slope will compensate with a stronger point source.  Relative to
\macsa, in which MUSTANG does not detect AGN emission, this effect
reduces the constraining power on $\gamma$, which can be seen by
comparing Figures~\ref{fig:m0647gamma} \&
\ref{fig:m1206gamma}. Measurements of the point source flux closer in
frequency to 90~GHz are required to model and remove the source prior
to fitting and thereby improve the constraining power on $\gamma$.

Previous analyses of \macsb\ suggest that the system is close to being
in dynamical equilibrium. \citet{gilmour2009} classify the cluster as
visually relaxed based on its X-ray morphology. The mass profiles
derived from galaxy kinematics \citep{biviano2013}, X-ray surface brightness,
and combined strong and weak-lensing \citep{umetsu2012} are all consistent, 
which indicates that the system is likely relaxed.

As described in \S\ref{sec:m1206}, MUSTANG detects an excess residual
of SZE flux ($>$3-$\sigma$) to the NE of the bulk ICM in \macsb, after
removing the point source and G7 SZE models. This signal does not
appear to have a counterpart in the X-ray surface brightness image,
nor is there a diffuse radio feature in GMRT observations that would
point to a shock associated with an energetic merger event
\citep[e.g.,][]{ferrari2011}.  When comparing the MUSTANG map to the
optical image and a weak lensing mass reconstruction using data and
methods presented in \citet{umetsu2012}, we do however see some
evidence that this source is aligned with a filamentary structure to
the N-NE (Figure~\ref{fig:m1206wl}).

\input{fig13.tex}

Figure~\ref{fig:m1206wl} shows an optical image of \macsb\ with weak
lensing mass contours overlaid. The SE elongation in the mass
distribution follows a filamentary structure that has been noted in
previous analyses \citep[see][]{umetsu2012,annunziatella2014}. Additionally,
there appears to be an elongation in the mass distribution to the NE,
in the direction of the feature detected by MUSTANG.  The centroid of
the SZE signal measured by Bolocam is also shifted to the NE (see
Figure~\ref{fig:m1206composite}).

\subsubsection{Galaxy Group Scenario}\label{sec:group}

We consider the case of a galaxy group leading to the SZE feature
detected by MUSTANG. Using the residual flux measured in the filtered,
model-subtracted MUSTANG SZE maps, we can place constraints on the
group mass.  By simulating a suite of idealized A10 cluster
Compton-$y$ maps, we find the MUSTANG residual can only provide a
lower limit to the mass, since the filtering effects remove an unknown
and possibly large SZE flux component from angular scales inaccessible
to MUSTANG, while for a small enough group or cluster little flux is
filtered.  We use this mass to infer what the X-ray surface brightness
of the group would be, and determine if such a lower limit is
consistent the upper limit placed by X-ray.  The residual integrated
SZE flux of $-61$~$\mu$Jy corresponds to a mass lower limit of
$M_{500}>1.3\times 10^{13}$~\msun\ and soft (0.1$-$2.4~keV) X-ray
luminosity of $\Lx>7.99 \times 10^{43}$ erg s$^{-1}$ (see
Table~\ref{tbl:m1206fluxes}). In this calculation we have assumed the
$Y-M$ and $Y-\Lx$ scaling relations given in A10. We note that
\Ysz$D_{A}^{2}=9.53 \times 10^{-8}$~Mpc$^{2}$ is below the mass limit of
the sample used in A10, so this is an extrapolation.  

Using the spectroscopic redshifts of \citet{biviano2013}, which are
part of the ``CLASH-VLT'' VIsible MultiObject Spectrograph (VIMOS)
Large Programme and have been recently made publicly available, we
analyze galaxy structures outside the main cluster peak in the
redshift distribution, selecting galaxies corresponding to foreground
and background peaks. One of these redshift bins, at ${z\sim 0.42}$,
contains 13 galaxies that are located near the SZE peak. We take these
galaxies to be members of a potential group associated with the SZE
feature and compute the line of sight velocity dispersion, \sigV. For
this group we find $\avg{z}\sim 0.423$ and
$\sigV=650$~km~s$^{-1}$. Therefore, this is potentially either a
foreground group $\sim$100~Mpc in front of the cluster or a group
falling into the cluster with a rest frame velocity of $V_{\rm rf}\sim
3500$~km~s$^{-1}$ toward the observer.

Following the \sigV-$M_{200}$ relation of \citet{munari2013},
we compute a group mass of
$M_{200}=(2.4\pm1.5)\times10^{14}$~\msun\ within $R_{200}\sim1.1$~Mpc,
corresponding to $M_{500}\approx1.4\pm 0.9\times 10^{14}$~\msun\ and
$R_{500}\approx0.7^{+0.1}_{-0.2}$~Mpc for a typical scaling of 
$M_{500}\approx 0.6 \times M_{200}$ of an NFW mass profile.

\input{fig14.tex}

Figure~\ref{fig:phase} shows the projected phase space diagram for the
galaxies in this study including escape velocity curves for both the
primary cluster and the potential group. We compute the escape
velocities using an NFW mass density profile and the procedure of
\citet{denhartog1996}. The escape velocity curves for the cluster and
the group are centered on the BCG and the brightest group galaxy
(BGG), respectively.  The BGG is located at ($\alpha_{\rm J2000},
\delta_{\rm J2000})=(12^{\rm h}06^{\rm m}13.2^{\rm s},
-08\degree47$\amin$45$\asec), within the MUSTANG SZE residual region.
Figure~\ref{fig:phase} shows an additional spiral galaxy at
($\alpha_{\rm J2000}, \delta_{\rm J2000})=(12^{\rm h}06^{\rm
  m}13.3^{\rm s},-08\degree47$\amin$37$\asec) that is coincident with
an X-ray compact source and optically brighter than the BGG, but was
originally assigned to the main cluster. Moreover, the Peak+Gap method
\citep[see][]{fadda1996,biviano2013} used to assign member galaxies to
the main cluster computes a $37\%$ probability that this galaxy
belongs with the $z\sim0.42$ population instead, making it a likely
member of the putative group.

We also use the \chandra\ data to provide a consistency check on the
putative group's mass.  Since the X-ray centroid of \macsb\ is shifted
toward the east (see green `X' in Figure~\ref{fig:m1206composite}),
showing excess emission just south of the group's location on the sky,
it is difficult to disambiguate the group's X-ray flux from that of
\macsb.  Further, for masses consistent with the velocity dispersion 
mass estimate above, $R_{500}$ of the group lies entirely within that of
\macsb, so the X-ray background in this region is higher than it would
be in a similar observation of an isolated group.  Therefore, our 
estimate of the X-ray luminosity of the undetected group can only 
provide a weak upper limit of its mass.

Using an aperture corresponding to the $R_{500}$ of the optically identified 
group, the exposure-corrected 0.1--2.4~keV \chandra\ image of \macsb\ 
yields an X-ray flux of 7.8$^{+0.9}_{-1.6}\times10^{-13}~\rm
erg~s^{-1}~cm^{-2}$. This provides an upper limit on the soft
0.1--2.4~keV X-ray luminosity of $\Lx<$ 4.3--6.1$\times
10^{44}$~erg~s$^{-1}$ for an infalling group masked by the 
X-ray emission from \macsb, which we have attempted to subtract from 
the flux estimate in this region.  
Using the Malmquist Bias corrected $\Lx-M_{500}$ scaling
relation of \cite{pratt2009}, which are consistent the $\Ysz-M$ and
$\Ysz-\Lx$ scaling relations in A10, we place an upper limit of 
$M_{500}< (4.0$--$4.5)\times10^{14}$~\msun\ for the region selected by 
the MUSTANG and VLT data.

Finally, we use a multi-halo NFW fitting procedure
\citep[see][]{medezinski2013} to derive weak-lensing mass estimates
for the group by fitting two halos, namely the main cluster and the putative 
group, using the gravitational shear data presented in \citet{umetsu2014} (see
also \citealt{umetsu2012}). To do this, we construct a reduced-shear map on
a regular grid of 42~$\times$~42 independent cells, covering a
24~$\times$~24 arcminute$^{2}$ region centered on the BCG. We exclude
from our analysis the 2~$\times$~2 innermost cells lying in the
supercritical (strong-lensing) regime.

We describe the primary cluster as an elliptical NFW (eNFW; see
\citealt{umetsu2012}; \citealt{medezinski2013}) model with the
centroid fixed at the BCG, thus specified with four parameters,
namely, the halo mass ($M_{200}$), concentration ($c_{200}$),
ellipticity ($e=1-b/a$), and position angle of the major axis. We
assume uniform priors for the halo mass, $M_{200}>0$, and the
concentration, $3\leq c_{200}\leq6$, which is the range expected for
CLASH X-ray selected clusters \citep[see][]{meneghetti2014}. For the
group, we assume a spherical NFW model with the centroid fixed at
($\alpha_{\rm J2000},\delta_{\rm J2000})= (12^{\rm h}06^{\rm
  m}13.3^{\rm s},-08\degree47$\amin$37$\asec), and the redshift at
$z=0.423$. We assume a flat prior for the group halo mass,
$M_{200}<7.5\times10^{14}$~\msun, corresponding to the X-ray-derived
upper limit of
$M_{500}\sim4.5\times10^{14}$~\msun\ (Table~\ref{tbl:masses}), and adopt the $c$-$M$ relation
from \citet{bhattacharya2013}. We marginalize over the source redshift
uncertainty reported in Table~3 of \citet{umetsu2014}.

The resulting two halo model is shown in Figure~\ref{fig:m1206wl}
(blue contours).  
From the simultaneous two-component (eNFW+NFW) fit to the
2-dimensional reduced shear data, we determine a group mass of
$M_{200}=3.6\pm2.0\times10^{14}$~\msun, or
$M_{500}=2.2\pm1.2\times10^{14}$~\msun. In this two-halo fit, we find
the best value of the primary cluster mass to be
$M_{200}=9.4\pm3.1\times10^{14}$~\msun, or
$M_{500}=5.6\pm1.9\times10^{14}$~\msun. We note that these results are
sensitive to the assumed priors as the weak-lensing data do not
resolve the group.

The X-ray and SZE measurements place constraints on the group mass
that, while not stringent, are consistent with the mass estimates from
the velocity dispersion and the multi-halo eNFW+NFW fitting (see
Table~\ref{tbl:masses}).

\input{tab7.tex}

\subsubsection{Extended Radio Emission}\label{sec:extradio}

GMRT observations at 610~MHz reveal extended diffuse radio emission
west of the central AGN (Figures~\ref{fig:m1206composite} \&
\ref{fig:m1206snr}), which is likely an AGN-driven plasma bubble or
jet.  However, such lobes are generally produced as symmetric pairs
powered by the central black hole.  The middle panel of
Figure~\ref{fig:m1206snr} shows that, after point source subtraction
to account for the AGN, there is an excess amount of pressure east of 
the diffuse radio emission.  This excess is associated with the core of
the best-fit gNFW model (Table~\ref{tbl:nfwmods})
that describes the MUSTANG and Bolocam data (yellow `X' in 
Figure~\ref{fig:m1206composite}). 
We note this model has a steeper inner profile than the median A10 
universal pressure profile.
We posit that this positional offset leads to suppression of
the would-be eastern radio lobe, while the western lobe appears be
expanding asymmetrically away from this higher pressure region.

The offset between the pressure profile and the AGN/BCG seems to suggest 
the ICM is sloshing subsonically, as sloshing should not produce strong 
pressure discontinuities \citep{zuhone2013}.
This scenario is supported by the $\sim$7\asec\ offset between the 
centroid of the X-ray emitting gas and the BCG location 
(Figure~\ref{fig:m1206composite}), along with
the general E-W elongation of both the surface mass distribution seen
in strong-lensing and the X-ray surface brightness.

The strong-lensing data in \citet{zitrin2012} (reproduced in
Figure~\ref{fig:m1206composite}) reveal a massive component
$\sim$40\asec\ to the east of the cluster core.  If this subcluster
has passed in front of or behind the main cluster's core, it may have
induced E-W sloshing that has redirected one or both jets away from
the line of sight.  In this case, the radio emission observed could be
a superposition of both lobes, or one lobe could be masked by the bright
AGN emission.  Sloshing also allows for the possibility that a detached,
aged lobe or bubble was compressed adiabatically, re-accelerating its
relativistic electrons to emit in the radio
\citep[e.g.,][]{Clarke2013}.  Deeper multi-band radio data are required
to measure the spectral index of the diffuse emission to distinguish
the possibilities. In addition, higher resolution radio data are
necessary to understand the nature of the western radio feature and
its interaction with the surrounding ICM and connection to, or
detachment from, the AGN.

In the sloshing scenario outlined above, the MUSTANG SZE residual 
substructure (right panel of Figure~\ref{fig:m1206snr}) is most 
plausibly an interloping foreground structure associated with the 
group discussed in Section~\ref{sec:group}. 

\section{CONCLUSION} 
\label{sec:conclusion}
We have presented high-resolution images of the SZE from MUSTANG
observations of \macsa\ and \macsb. We compare the MUSTANG
measurements to cluster profiles derived from fits to lower resolution
Bolocam SZE data and find that in general a steeper core profile is
called for compared to the universal pressure profile from A10. We
note that this fitting procedure does not perform a true simultaneous
fit to both the MUSTANG and the Bolocam data, which will be presented
in \romb.

We use archival \chandra\ data to generate pseudo-SZE models for
both \macsa\ and \macsb, which we normalize based on the integrated
flux within a 1\amin\ radius from the Bolocam observations. We find
that the Bolocam SZE profile in the core is shallower than the
pseudo-SZE, which we attribute to smoothing by the Bolocam PSF on the
scales shown in these maps.

In \macsa\ the MUSTANG SZE decrement closely follows the shape and
flux expected from the X-ray pseudo-SZE map. The MUSTANG and Bolocam
data are well described by a gNFW model with $\gamma=0.9$. MUSTANG
does not find any strong evidence for departures from hydrostatic
equilibrium in \macsa.

MUSTANG detects the central AGN in \macsb\ in addition to an excess of
SZE emission to the NE. We compare the MUSTANG data to models derived
from Bolocam and find that a gNFW with $\gamma=0.7$ best describes the
data. After accounting for the point source and primary ICM
distribution, MUSTANG measures a $\sim$3-$\sigma$ residual decrement
to the NE. Using spectroscopic redshift measurements, we carry out a
kinematic analysis of the galaxies surrounding the main cluster and
find evidence for a 13 member group at $z\sim0.42$. From the X-ray and
SZE data we derive upper and lower bounds, respectively, for the mass
of this group. We carry out a multi-halo fit to constrain a
weak-lensing mass estimate for the group and find good agreement with
the mass derived from the VLT data.

Observations with the GMRT at 610~MHz reveal extended radio emission
west of the central AGN. We suggest that this emission is an
AGN-driven plasma bubble or jet. While deeper multi-wavelength and
higher resolution data are required to characterize this feature, the
asymmetric morphology of the proposed jet could be explained by
sloshing of the ICM or an infalling group to the NE.


\acknowledgements The National Radio Astronomy Observatory is a
facility of the National Science Foundation operated under cooperative
agreement by Associated Universities, Inc. The GBT+MUSTANG
observations presented here were obtained with telescope time
allocated under NRAO proposal IDs AGBT11A009, and AGBT11B001.  

The Bolocam observations presented here were obtained operating from
the Caltech Submillimeter Observatory, which, when the data used in
this analysis were taken, was operated by the California Institute of
Technology under cooperative agreement with the National Science
Foundation. Bolocam was constructed and commissioned using funds from
NSF/AST-9618798, NSF/AST-0098737, NSF/AST-9980846, NSF/AST-0229008,
and NSF/AST-0206158. Bolocam observations were partially supported by
the Gordon and Betty Moore Foundation, the Jet Propulsion Laboratory
Research and Technology Development Program, as well as the National
Science Council of Taiwan grant NSC100-2112-M-001-008-MY3.

The \macsb\ spectroscopic data were based on the ESO VLT Large
Programme (prog.ID 186.A-0798, PI: P. Rosati).

Basic research in radio astronomy at the Naval Research Laboratory is
supported by 6.1 Base funding.

The late night assistance of the GBT operators Greg Monk, Donna
Stricklin, Barry Sharp and Dave Rose was much appreciated during the
observations.  Much of the work presented here was supported by NSF
grant AST-0607654.  Support for TM was provided by NASA through the
Einstein Fellowship Program, grant PF0-110077, and through a National
Research Council Research Associateship Award at the U.S.\ Naval
Research Laboratory.  Support for PK and AY was provided by the NRAO
Student Observing Support (SOS) and NASA Postdoctoral Fellowship
programs. JS was partially supported by a Norris Foundation CCAT
Postdoctoral Program Fellowship and by NSF/AST-1313447.


\input{young_mustang.bbl}
\end{document}

%% file: tab1.tex
\begin{deluxetable*}{lcccccc}
\tabletypesize{\scriptsize}
\tablewidth{0pt}
\tablecolumns{6}
\tablecaption{Cluster Properties}
\tablehead{
Cluster	 & $z$ &  $R_{500}$ & $P_{500}$ & $M_{500}$ & \kB\Tx & $Y_{500}$\tablenotemark{a} \\
         &  & (Mpc)    & $(10^{-3}$~keV cm$^{-3})$    & $(10^{14}$~\msun)  & (keV) & $(10^{-10}$~sr$)$ 
} 
\startdata 
MACSJ0647.7 & 0.591  & $1.26\pm 0.06$	& $9.23\pm 2.57$     & $10.9\pm 1.6$  & $11.5\pm 1.1$  & $1.7\pm 0.5$ \\[0.25pc]
MACSJ1206.2 & 0.439  & $1.61\pm 0.08$	& $10.59\pm 3.07$    & $19.2\pm 3.0$  & $10.7\pm 1.3$  & $5.5\pm 1.6$
\enddata
\label{tbl:clusprops}
\tablenotetext{a}{Values of $Y_{500}$ are derived from the values reported in \citealt{mantz2010b} 
using Equation~\ref{eq:y500}. }
\tablecomments{X-ray--derived cluster properties, reproduced from \citealt{mantz2010b}.}
\end{deluxetable*}

%% file: tab2.tex
\begin{deluxetable}{lcccc}
\tabletypesize{\scriptsize}
\tablewidth{0pt}
\tablecolumns{5}
\tablecaption{MUSTANG Observation Overview}
\tablehead{
Cluster	 & \multicolumn{2}{c}{Centroid (J2000)}   & Obs. Time & Peak $|$S/N$|$ \\
        		&  R.A. 	& Dec. 	  & (hrs)  & 
} 
\startdata 
MACSJ0647.7       & 06:47:50.5 	& $+$70:14:53 	& 16.4 & 8.1 \\[.25pc]
MACSJ1206.2       & 12:06:12.5	& $-$08:48:07 	& 12.1 & 4.1
\enddata
\label{tbl:obs}
\tablecomments{MUSTANG observations were carried out between February
  2011 and January 2013.}
\end{deluxetable}

%% file: fig1.tex
\begin{figure*}[tbh!]
  \centerline{
    \includegraphics[height=2.3in]{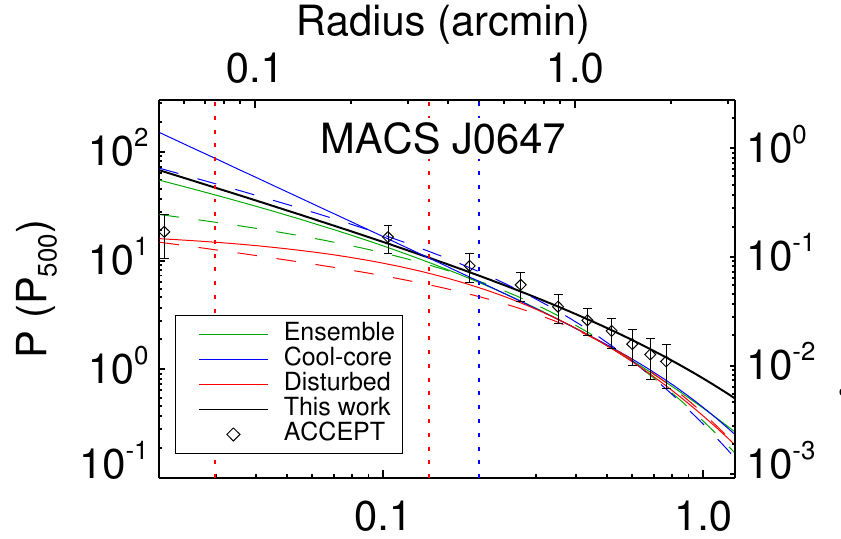}
    \includegraphics[height=2.3in]{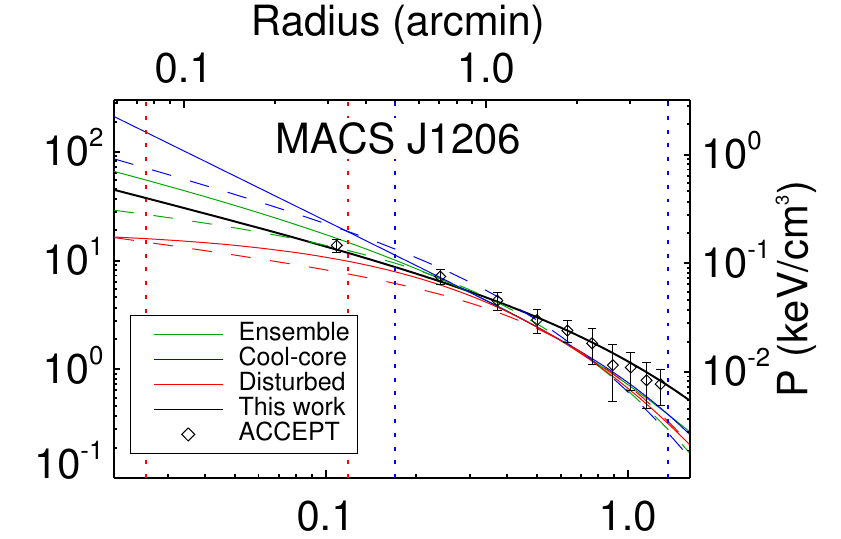}
  }
  \centerline{
    \includegraphics[height=2.3in]{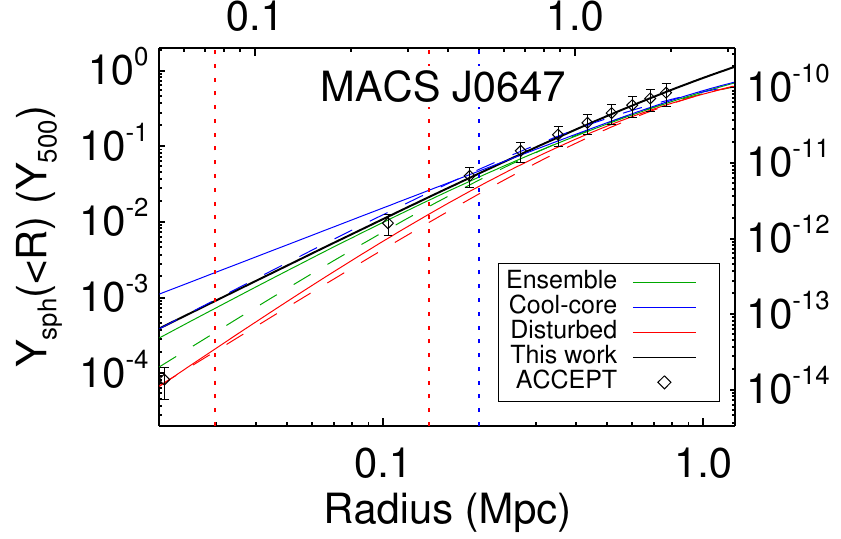}
    \includegraphics[height=2.3in]{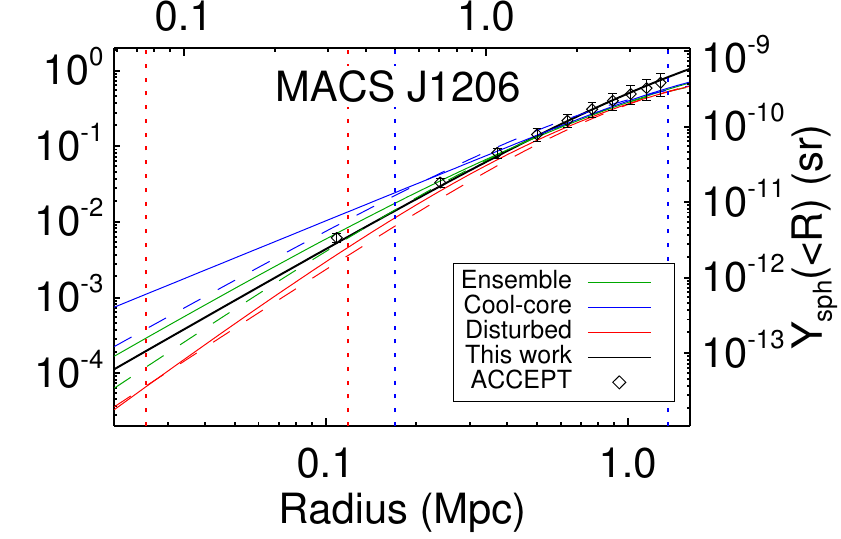}
  }
  \caption{\footnotesize Pressure (upper) and spherically-integrated
    Compton-$y$ (lower) profiles for \macsa\ (left) and
    \macsb\ (right). Dashed lines refer to the A10 sample of X-ray
    selected clusters, while solid lines correspond to the S13 sample
    including all of the CLASH clusters. For A10 and S13,
    respectively, ``ensemble'' (green) refers to the entire cluster
    sample, and profiles for cool-core (blue) and disturbed (red)
    morphologies are also separately shown.  The X-ray derived
    pressure measurements from the ACCEPT database are plotted as
    diamonds. The best-fit Bolo+MUSTANG model presented in
    \S\ref{sec:results} is given by the solid black line in each
    plot. The vertical dotted lines surround the radial dynamic range
    (resolution to FOV) covered by MUSTANG (red) and Bolocam
    (blue). Note the Bolocam FOV extends beyond the radial range shown
    for \macsa. The integrated Compton-y profiles were computed
    according to Equations~\ref{eq:ysph}~\&~\ref{eq:y500}.
    \label{fig:profiles}}
\end{figure*}

%% file: tab3.tex
\begin{deluxetable}{lcccccc}
\tabletypesize{\scriptsize}
\tablecolumns{7}
\tablecaption{gNFW Model Parameters}
\tablehead{
Model	           & $P_{0}$ & $C_{500}$ & $\gamma$ & $\alpha$ & $\beta$ \\
} 
\startdata 
S13 Ensemble    &   4.29  &  1.18   &  0.67     & 0.86    & 3.67    \\[.25pc]
S13 Cool-core   &   0.65  &  1.18   &  1.37     & 2.79    & 3.51    \\[.25pc]   
S13 Disturbed   &   17.3  &  1.18   &  0.02     & 0.90    & 5.22    \\[.25pc]   
A10 Ensemble    &   8.40  &  1.18   &  0.31     & 1.05    & 5.49    \\[.25pc]  
A10 Cool-core   &   3.25  &  1.13   &  0.77     & 1.22    & 5.49    \\[.25pc]  
A10 Disturbed   &   3.20  &  1.08   &  0.38     & 1.41    & 5.49    \\[.25pc] 
MACS J0647.7    &   0.54  &  0.29   &  0.90     & 1.05    & 5.49    \\[.25pc] 
MACS J1206.2    &   1.13  &  0.41   &  0.70     & 1.05    & 5.49    \\
\enddata
\label{tbl:nfwmods}
\vspace{-0.3cm} 
\tablecomments{Best-fit gNFW models from S13, A10, and the best-fit
  Bolo+MUSTANG models presented in \S\ref{sec:results}.} 
\end{deluxetable}

%% file: fig2.tex
\begin{figure*}[ht!]
  \centerline{
    \includegraphics[height=2.3in]{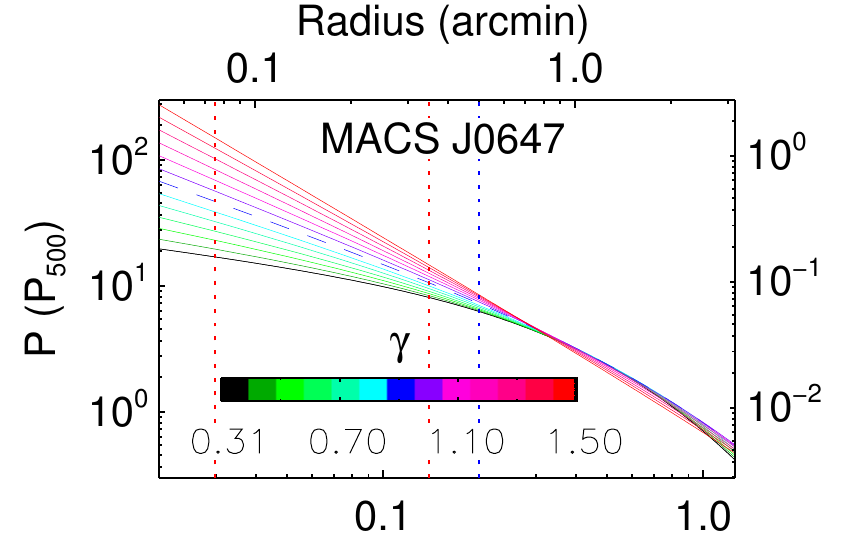}
    \includegraphics[height=2.3in]{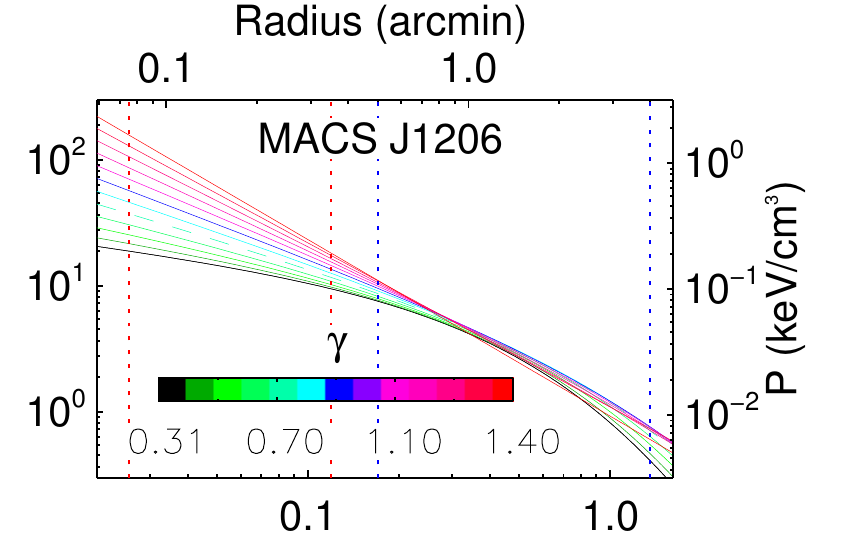}
  }
  \centerline{
    \includegraphics[height=2.3in]{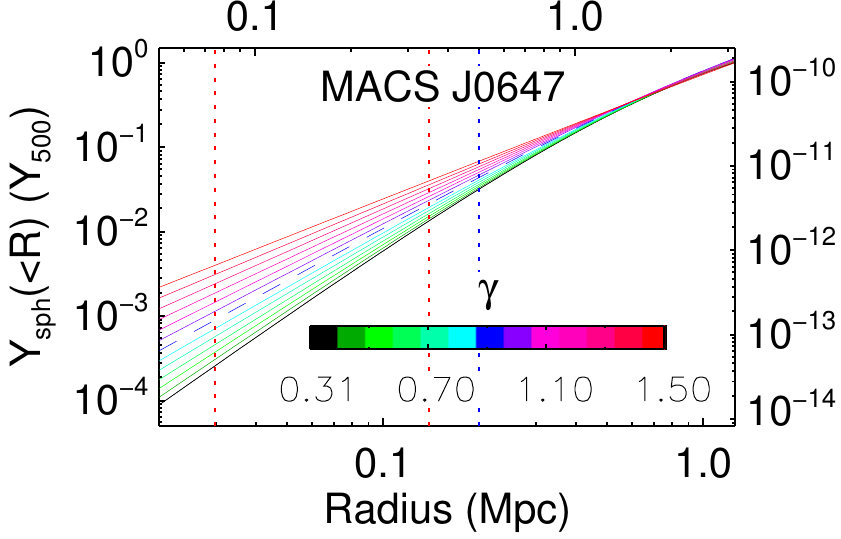}
    \includegraphics[height=2.3in]{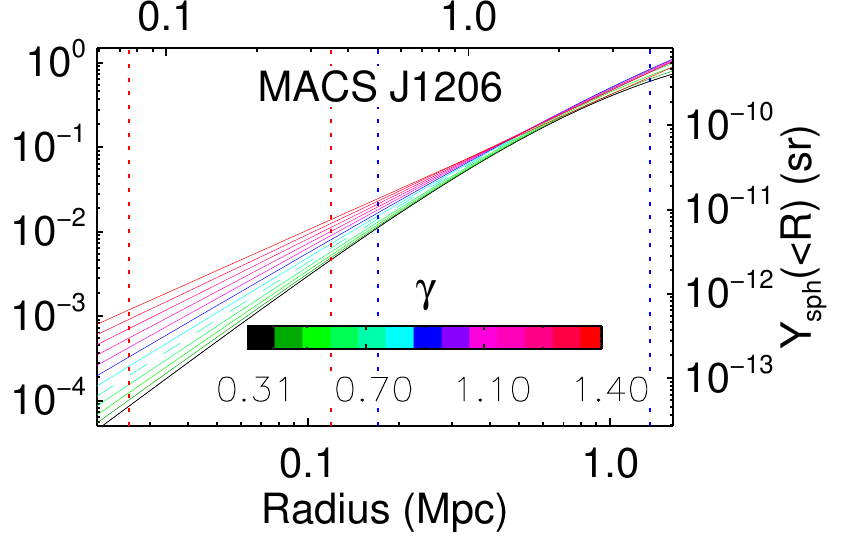}
  }
  \caption{\footnotesize Pressure (upper) and spherically-integrated Compton-$y$
    (lower) profiles generated from fits of generalized NFW profiles
    to Bolocam measurements of \macsa\ and \macsb, in the 
    left and right columns, respectively. Each profile
    represents the gNFW that best fits the Bolocam data given a fixed
    value of $\gamma$, represented by the color bars, with $\alpha$
    and $\beta$ held at the A10 values. In general, Bolocam has the
    largest constraining power between 1\amin\ and 3\amin\ in radius,
    and all of the models overlap to a high degree in this radial
    region. This highlights the inherent parameter degeneracies
    between $P_{0}$, $C_{500}$, and $\gamma$ in the gNFW model, which
    can be broken using the high-resolution MUSTANG data. The dashed
    lines correspond to the best fit Bolo+MUSTANG models, which have 
    $\gamma=0.9$ for \macsa, and $\gamma=0.7$ for \macsb\ 
    (see Table \ref{tbl:nfwmods}). 
    From left to right, the vertical dotted lines
    mark the resolution and FOV, respectively, of MUSTANG (red) and
    Bolocam (blue), as in Figure~\ref{fig:profiles}.
    \label{fig:bolomodels}}
\end{figure*}

%% file: fig3.tex
\begin{figure}[ht!]
\begin{center}
    \includegraphics[width=3.25in]{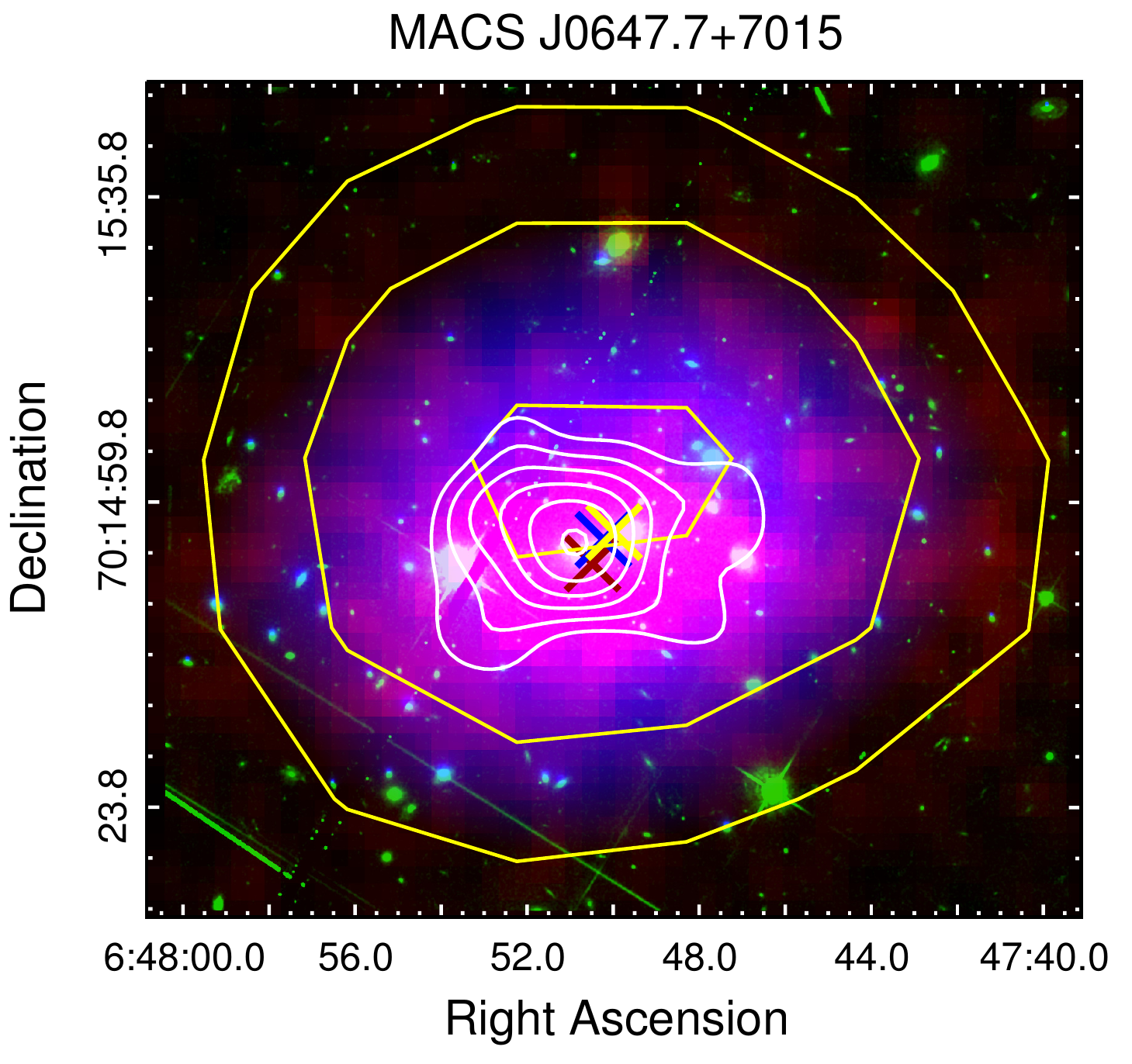}
\end{center}
  \caption{\footnotesize Composite image of \macsa. Green is HST, blue is the total
    mass distribution derived from strong gravitational lensing
    \citep{zitrin2011}, and red is X-ray surface brightness measured
    by \chandra.  MUSTANG S/N contours from Figure~\ref{fig:m0647snr}
    are overlaid in white and Bolocam contours (arbitrary units) are
    overlaid in yellow. Although the Bolocam peak is located slightly
    north of the cluster center, there is good agreement in general
    between the X-ray, SZE, and lensing mass distributions. Crosses
    denote the centroid for the X-ray surface brightness (dark red),
    BCGs (blue), and Bolocam SZE (yellow). \macsa\ exhibits an
    elliptical morphology with two distinct cD galaxies, which may
    indicate merger activity, but otherwise appears to be relaxed. The
    blue cross above is centered between the two cD galaxies.
    \label{fig:m0647composite}}
\end{figure}

%% file: fig4.tex
\begin{figure}[ht!]
  \begin{center}
    \includegraphics[width=3in]{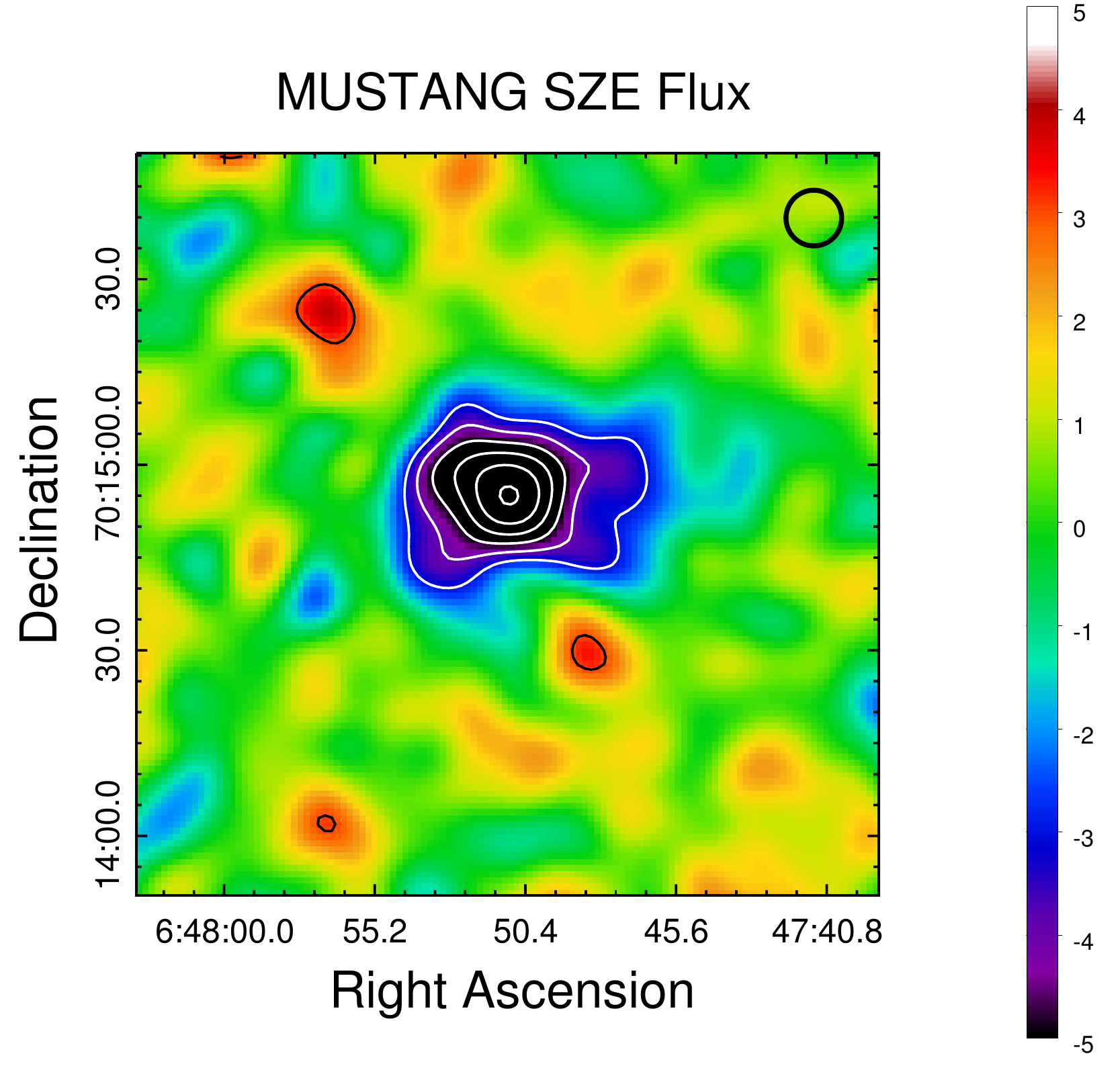}
  \end{center}
  \caption{\footnotesize MUSTANG SZE S/N map of \macsa\ smoothed with the
    9\asec\ beam represented by the black circle in the upper right.
    Contours are shown in increments of $1$-$\sigma$ beginning at
    $3$-$\sigma$ for SZE decrement (white) and positive flux (black).
    \label{fig:m0647snr}}
\end{figure}

%% file: fig5.tex
\begin{figure}[htb!]
  \begin{center}
    \includegraphics[width=3.5in]{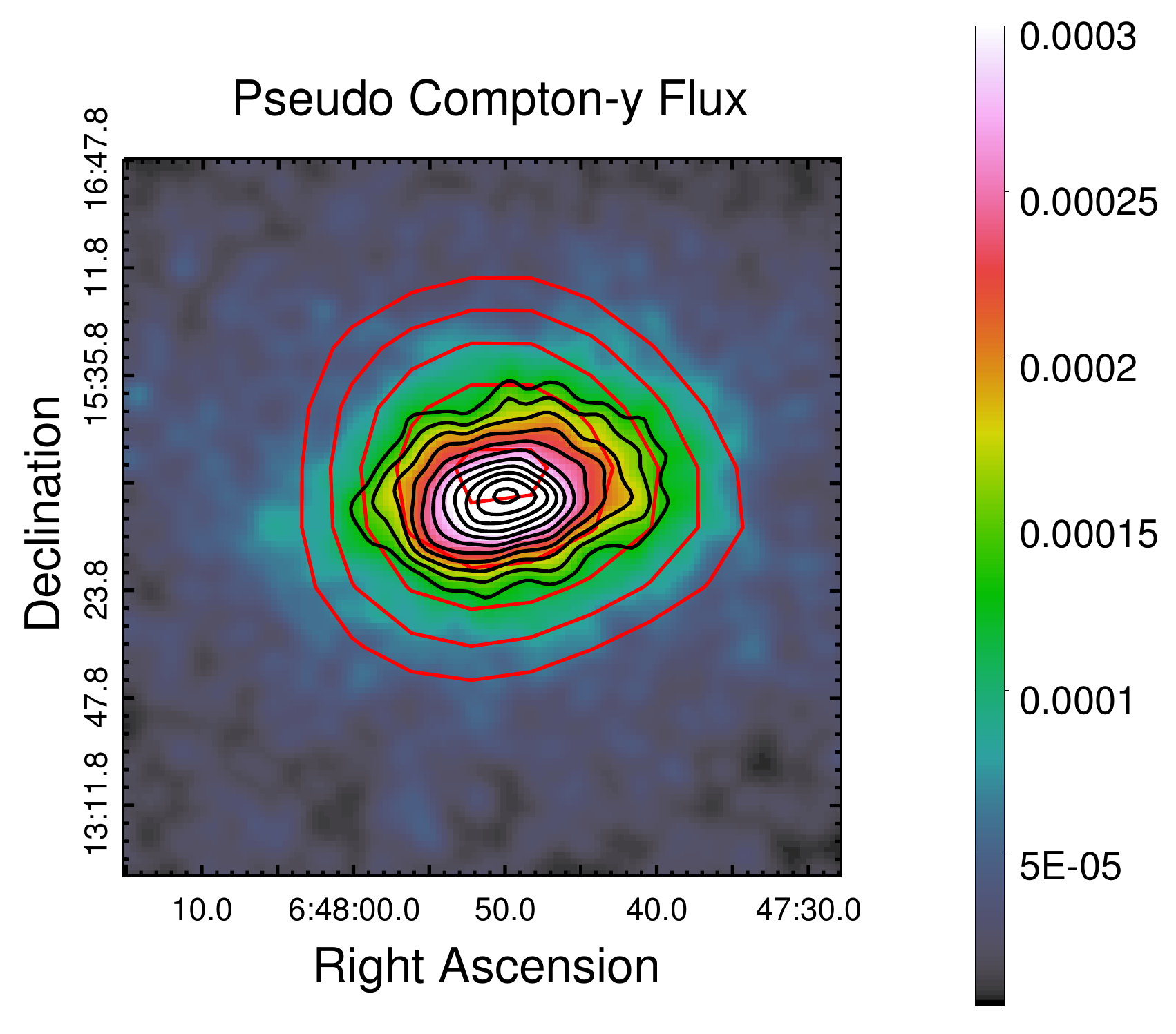}
\end{center}
  \caption{\footnotesize \macsa\ X-ray derived Compton-y map assuming an isothermal
    temperature of 11.5 keV and effective depth $\ell= 1.4$~Mpc. The
    contours are shown for X-ray \mbox{pseudo-$y$} (black) and Bolocam
    data (red) in increments of $2.6 \times 10^{-5}$ beginning at
    $1.3\times 10^{-4}$ for both. The Bolocam PSF smooths the signal
    significantly on the scale of this image, which explains the
    broader contours relative to the X-ray.
    \label{fig:m0647xray}}
\end{figure}

%% file: fig6.tex
\begin{figure}[ht!]
  \begin{center}
    \includegraphics[width=3in]{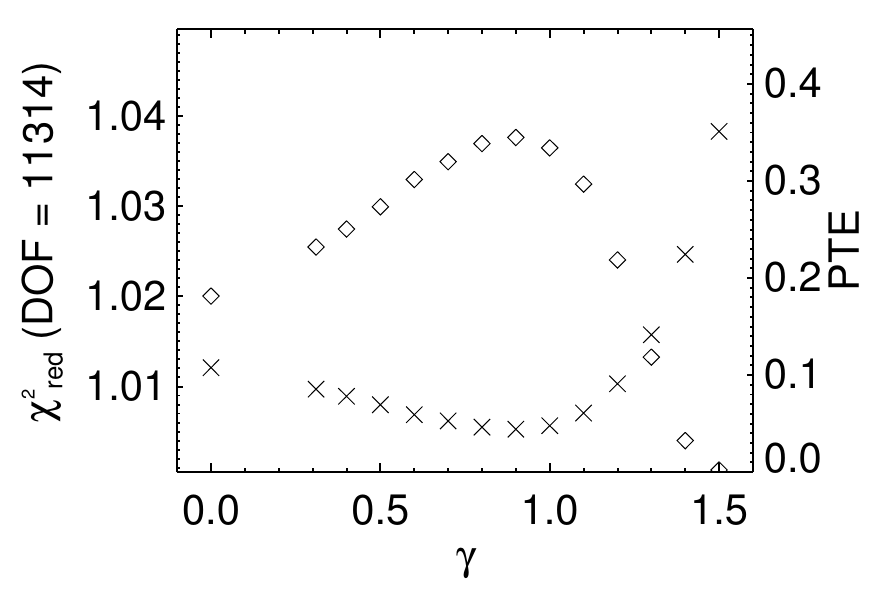}
  \end{center}
  \caption{\footnotesize Goodness of fit parameters
    \chired\ (crosses) and PTE (diamonds) from the comparison between
    MUSTANG data and the Bolocam-derived models for \macsa. We
    determine the best-fit model to be a gNFW with $\gamma$ =
    0.90, yielding $\chi^{2}/$DOF~$=11374/11314$ and PTE = 0.34.
    \label{fig:m0647gamma}}
\end{figure}

%% file: fig7.tex
\begin{figure}[tb]
  \includegraphics[width=3.3in]{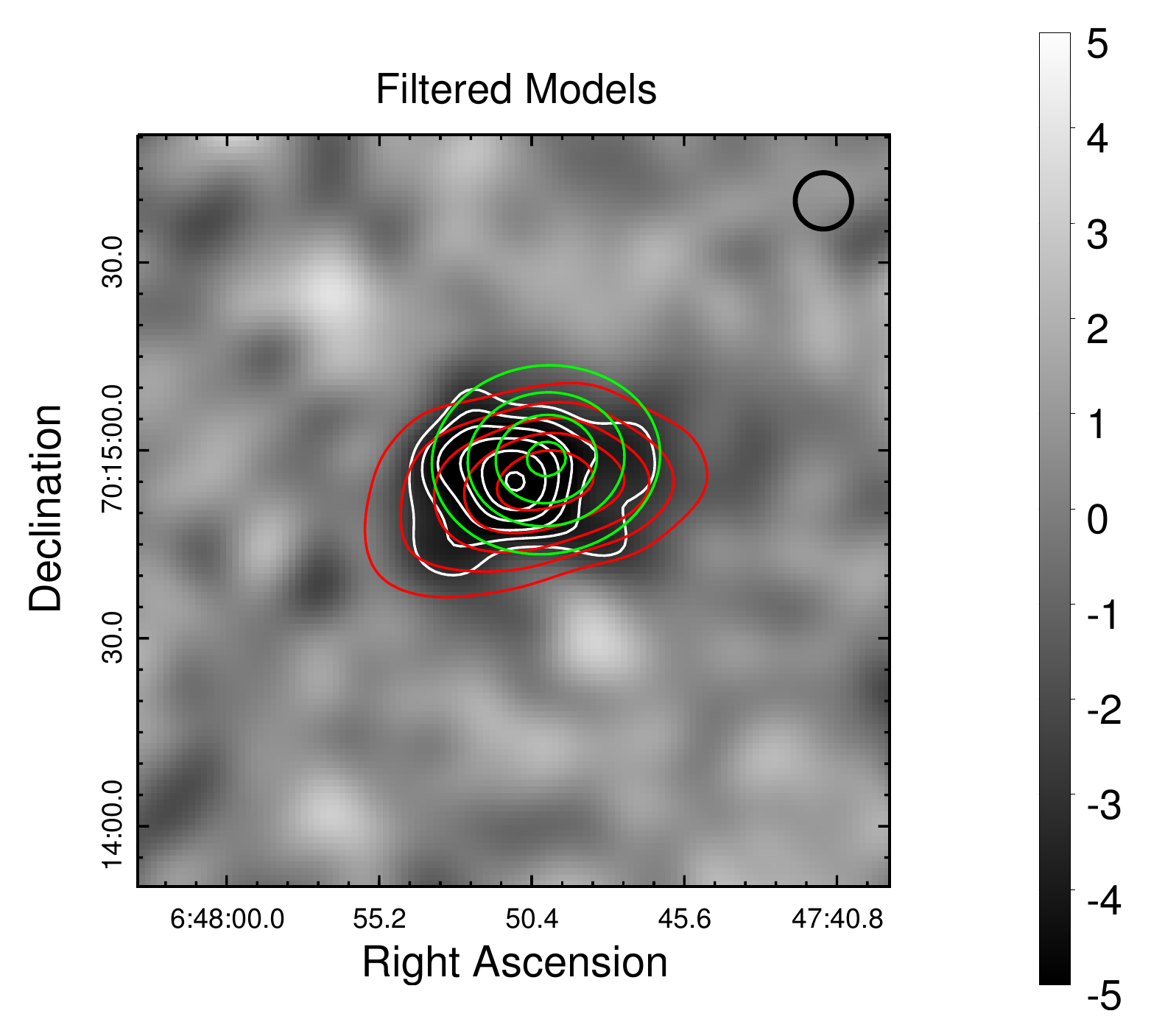}\\
  \includegraphics[width=3.1in]{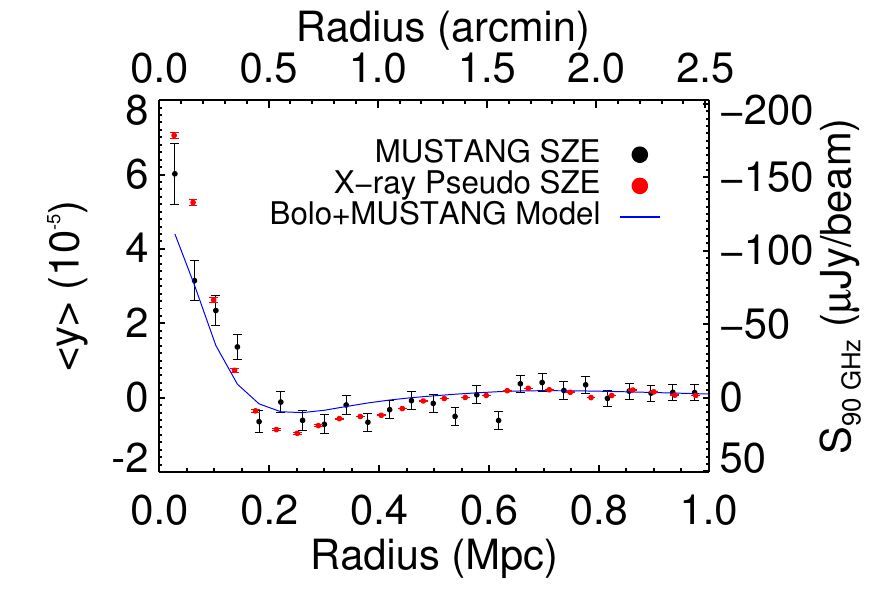}
  \caption{\footnotesize \macsa\ pseudo SZE map derived from
    \chandra\ X-ray data (top, red contours) with green contours
    representing the Bolo+MUSTANG G9 model from this work. The MUSTANG
    transfer function has been applied to both and the MUSTANG PSF is
    shown as a black circle. The white contours are MUSTANG S/N from
    Figure~\ref{fig:m0647snr}. The red and green contours are overlaid
    in units of -50~$\mu$Jy~beam$^{-1}$ starting at
    -50~$\mu$Jy~beam$^{-1}$. Azimuthally averaged radial profiles are shown
    in the lower panel. Aside from the central $\sim$0.1~Mpc
    where the X-ray and SZE flux are sharply peaked, the radially
    averaged flux from MUSTANG closely follows both the G9 model and
    the X-ray pseudo SZE flux.
    \label{fig:m0647inmod}}
\end{figure}

%% file: tab4.tex
\begin{deluxetable}{llcc}
\tablewidth{0pt}
\tabletypesize{\scriptsize}
\tablecolumns{4}
\tablecaption{Summary of Fit Results\label{tbl:fitresults}}
\tablehead{
Cluster &	Model &  \chisq/DOF  &  PTE 
}
\startdata
{\macsa} & & &\\[.05pc]
\hline \\[-0.5pc]
      &  A10              & $11425/11314$ & $0.23$ \\[.25pc]
      &  G9               & $11378/11314$ & $0.34$  \\[.25pc]
      &  Pseudo-SZE          & $11497/11314$ & $0.11$  \\[.1pc]
\hline \\[-.25pc]
{\macsb} & & &\\[.05pc]
\hline \\[-0.5pc]
      &  A10              & $11237/11307$ & $0.68$ \\[.25pc]
      &  G7               & $11227/11307$ & $0.70$  \\[.25pc]
      &  Pseudo-SZE          & $11408/11307$ & $0.25$ \\[-.7pc]
\enddata \tablecomments{Fit results for the A10, gNFW, and pseudo-SZE
  models. The fits for \macsb\ included a model for the point source
  with a floating amplitude.}
\end{deluxetable}

%% file: fig8.tex
\begin{figure}[tbh!]
\begin{center}
    \includegraphics[width=3.25in]{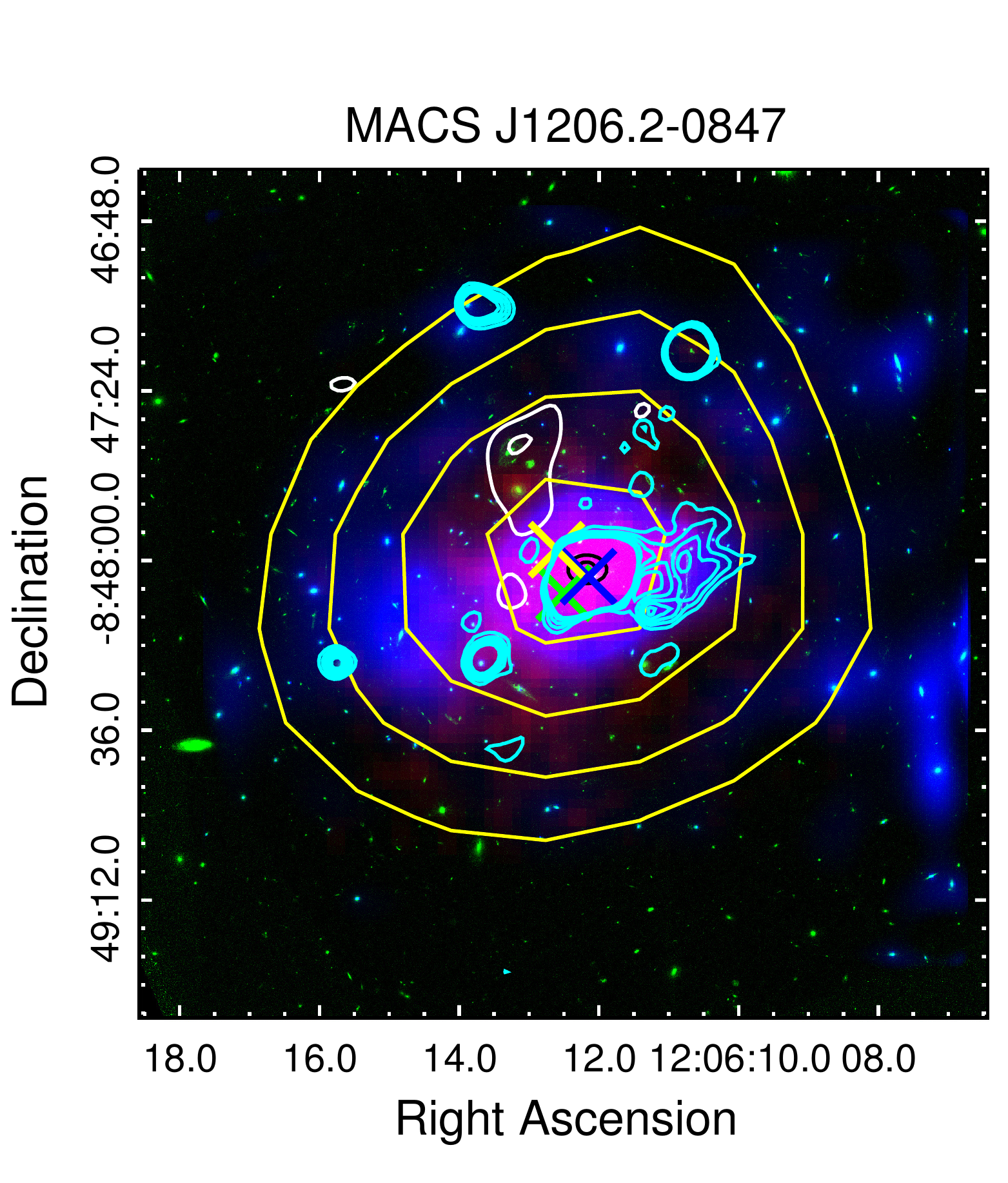}
\end{center}
  \caption{\footnotesize Composite image of \macsb.  Green is HST,
    blue is the total mass distribution derived from strong
    gravitational lensing \citep{zitrin2012}, and red is X-ray surface
    brightness measured by \chandra.  MUSTANG S/N contours from
    Figure~\ref{fig:m1206snr} are overlaid in white (negative) and
    black (positive). Bolocam contours (arbitrary units) are overlaid
    in yellow. Radio contours from GMRT 610~MHz observations are
    overlaid in cyan, and span 8-$\sigma$ to 17-$\sigma$ in steps of
    3-$\sigma$.  The crosses denote the centroids from the Bolocam
    data (yellow), the diffuse X-ray distribution (green), and the BCG
    (blue). The offsets between these centroids, as well as the
    extended radio emission, could be indicative of a disturbed
    cluster morphology.
    \label{fig:m1206composite}}
\end{figure}

%% file: fig9.tex
\begin{figure*}[tbh!]
  \centerline{
    \includegraphics[height=2.5in]{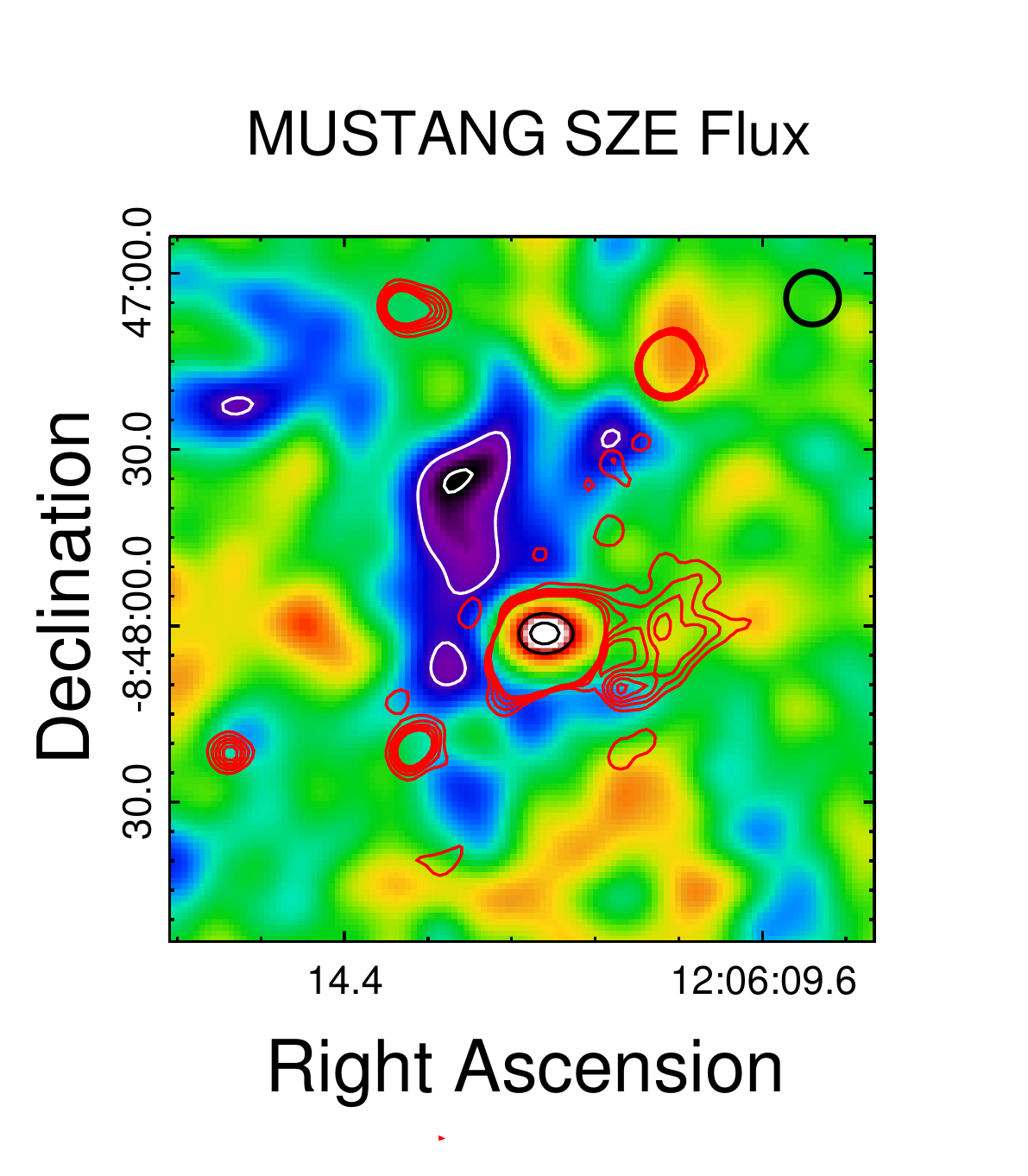}
    \includegraphics[height=2.5in]{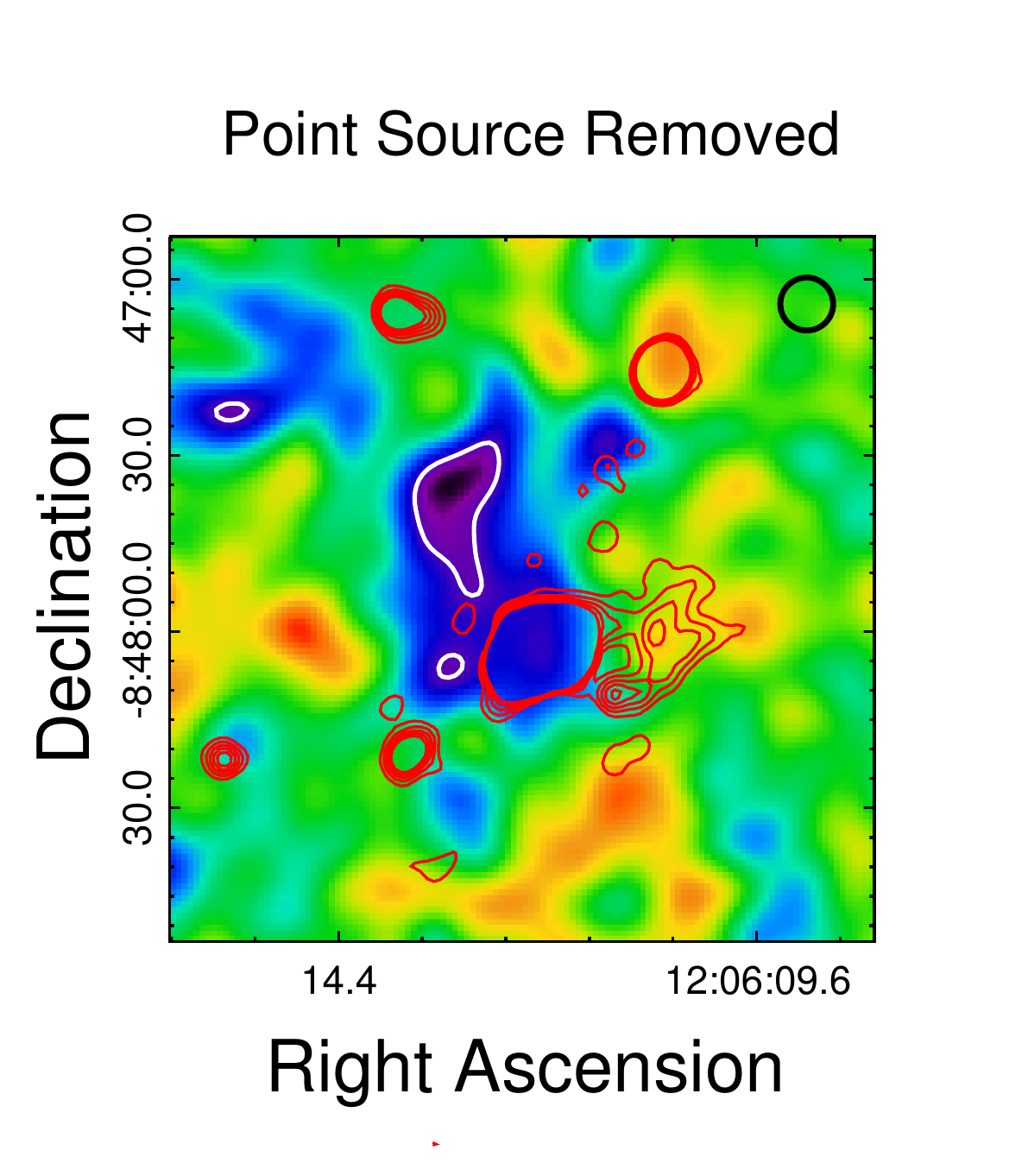}
    \includegraphics[height=2.5in]{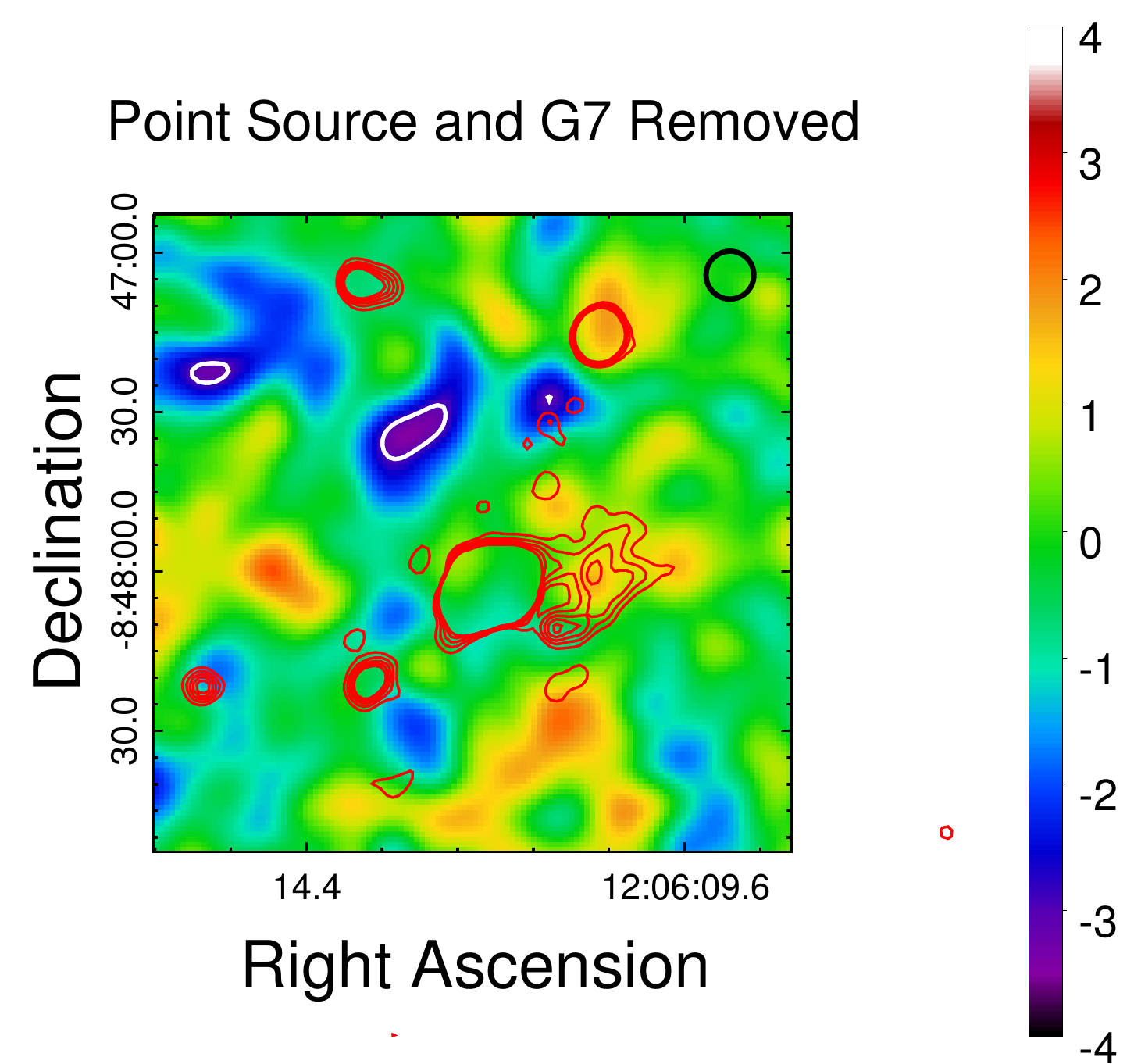}
  }
  \caption{\footnotesize MUSTANG S/N map of \macsb\ (left). Also shown
    are MUSTANG S/N maps with a point source model subtracted
    (middle), and additionally the G7 model subtracted (right). Black
    (white) contours are positive (negative) S/N = [3, 4]. There is a
    residual flux of $S_{90}=-$\macsbflux~$\mu$Jy to the NE with
    $>$3-$\sigma$ significance and not accounted for by the G7
    model. We include contours in red from GMRT observations at
    610~MHz, spanning 8-$\sigma$ to 17-$\sigma$ in steps of
    3-$\sigma$.  In each panel, the 9\asec\ MUSTANG beam is drawn as a
    black circle in the upper right.
    \label{fig:m1206snr}}
\end{figure*}

%% file: fig10.tex
\begin{figure}[t!]
  \begin{center}
    \includegraphics[width=3.5in]{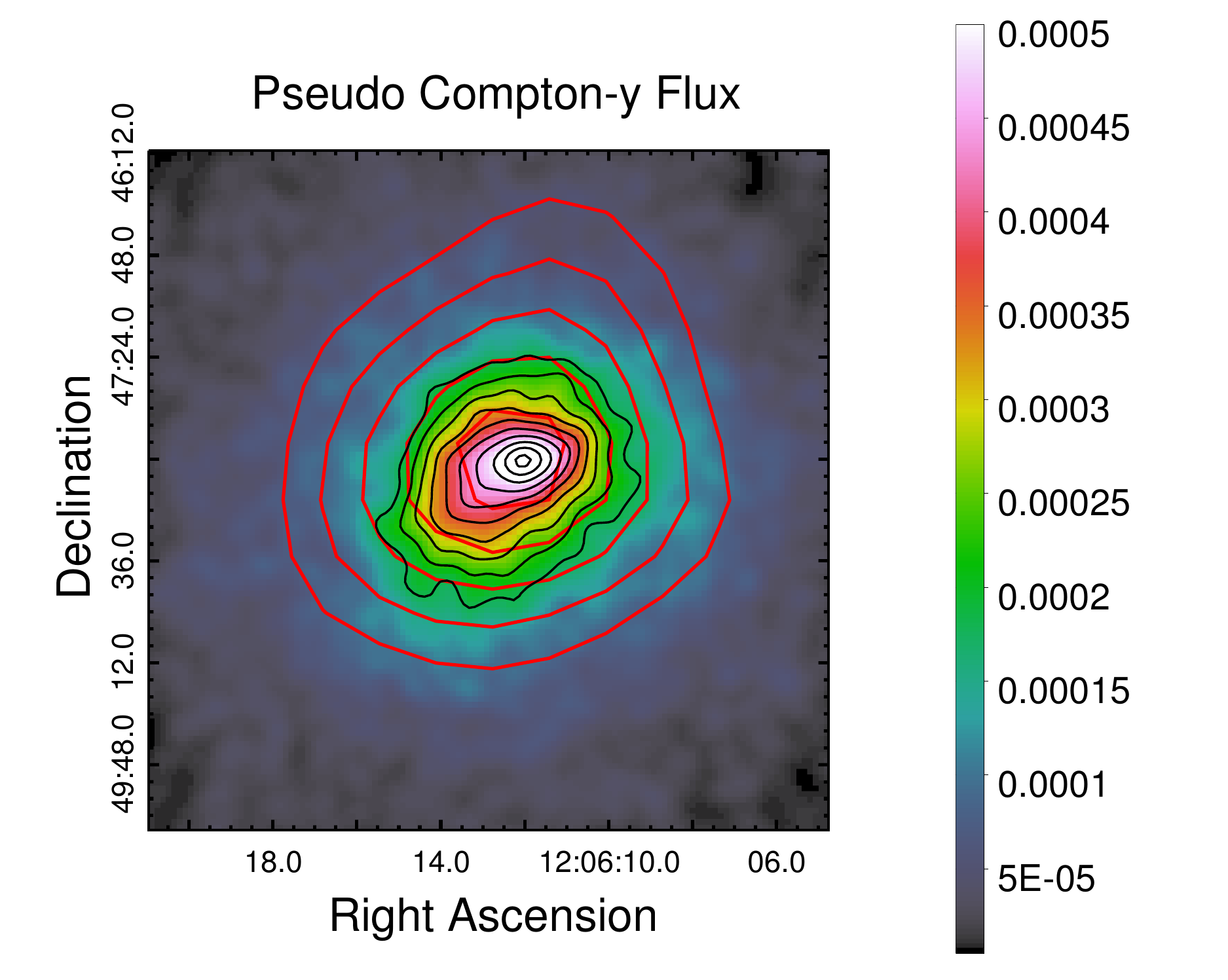}
  \end{center}
  \caption{\footnotesize \macsb\ X-ray derived Compton-y map assuming
    an isothermal temperature of 10.7~keV and effective depth $\ell\ =
    2.0$~Mpc. The contours are shown for X-ray \mbox{pseudo-$y$}
    (black) and Bolocam data (red) in increments of $0.25 \times
    10^{-4}$ beginning at $1.7\times 10^{-4}$ for both. As in
    Figure~\ref{fig:m0647xray}, the Bolocam contours are broader than
    the X-ray, due to the smoothing of the core flux by the Bolocam
    PSF.
    \label{fig:m1206xray}}

\end{figure}

%% file: tab5.tex
\begin{deluxetable}{lccc}
\tablecolumns{4} 
\tabletypesize{\scriptsize}
\tablecaption{Point Source Flux and Extrapolated Spectral Indices} 

\tablehead{ Model & $S_{90}$ & $\alpha$ & $\beta$ \\ & ($\mu$Jy) & & }

\startdata 
SPECFIND & $879\pm 253$ & $-1.26 \pm 0.09$ & $6.19 \pm 0.24$ \\[.25pc] 
A10 & $674\pm 61 $ & $-1.32\pm0.05$ & $6.34\pm0.25$ \\[.25pc] 
G7 & $765\pm 61 $ & $-1.28\pm0.05$ & $6.25\pm0.24$ \\[.25pc] 
Null & $584\pm 61 $ & $-1.35\pm0.05$ & $6.45\pm0.25$ 
\enddata

\tablecomments{Point source fluxes derived from joint fits with bulk
  SZE models. The first row provides the flux at 90~GHz (S$_{90}$)
  extrapolated from measurements at lower frequencies (74--1400~MHz)
  given in the SPECFIND V2.0 catalog \citep{vollmer2010}. The A10
  model refers to the ensemble parameters given in
  Table~\ref{tbl:nfwmods}. The G7 model is the best-fit Bolo+MUSTANG
  model from this work. The ``null'' model assumes there is no SZE
  decrement coincident with the point source. This represents a lower
  limit on the flux at 90 GHz and and therefore the steepest (most
  negative) likely spectral index.}
\label{tbl:ptsrc}
\end{deluxetable}

%% file: fig11.tex
\begin{figure}[ht!]
  \begin{center}
    \includegraphics[width=3in]{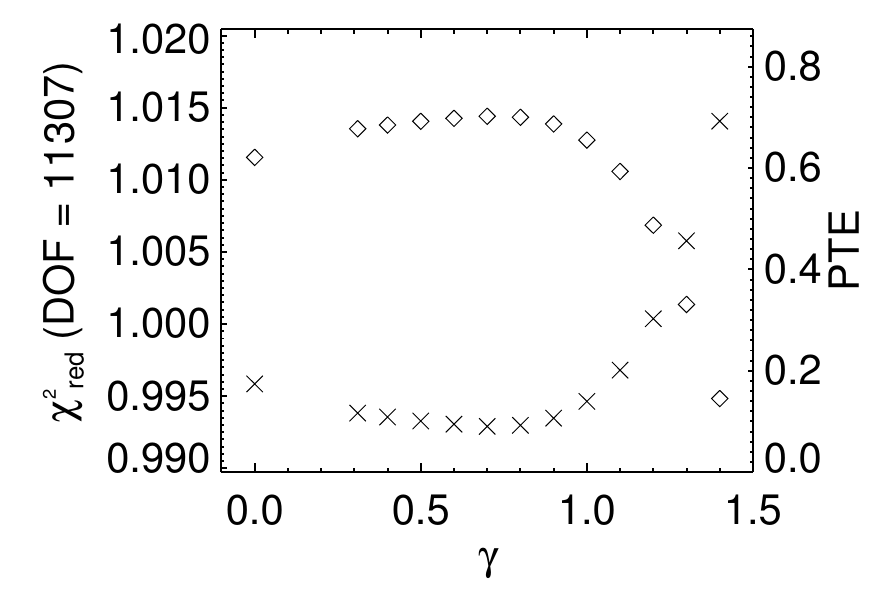}
  \end{center}
  \caption{\footnotesize Goodness of fit parameters
    \chired\ (crosses) and PTE (diamonds) from the comparison between
    MUSTANG data and the Bolocam-derived models for \macsb. We
    determine the best-fit model to be a gNFW ($\gamma$ = 0.70), for
    which we calculate $\chi^{2}/$DOF~$=11227/11307$ and
    PTE = 0.70.
    \label{fig:m1206gamma}}
\end{figure}

%% file: fig12.tex
\begin{figure}[tbh]
    \includegraphics[width=3.3in]{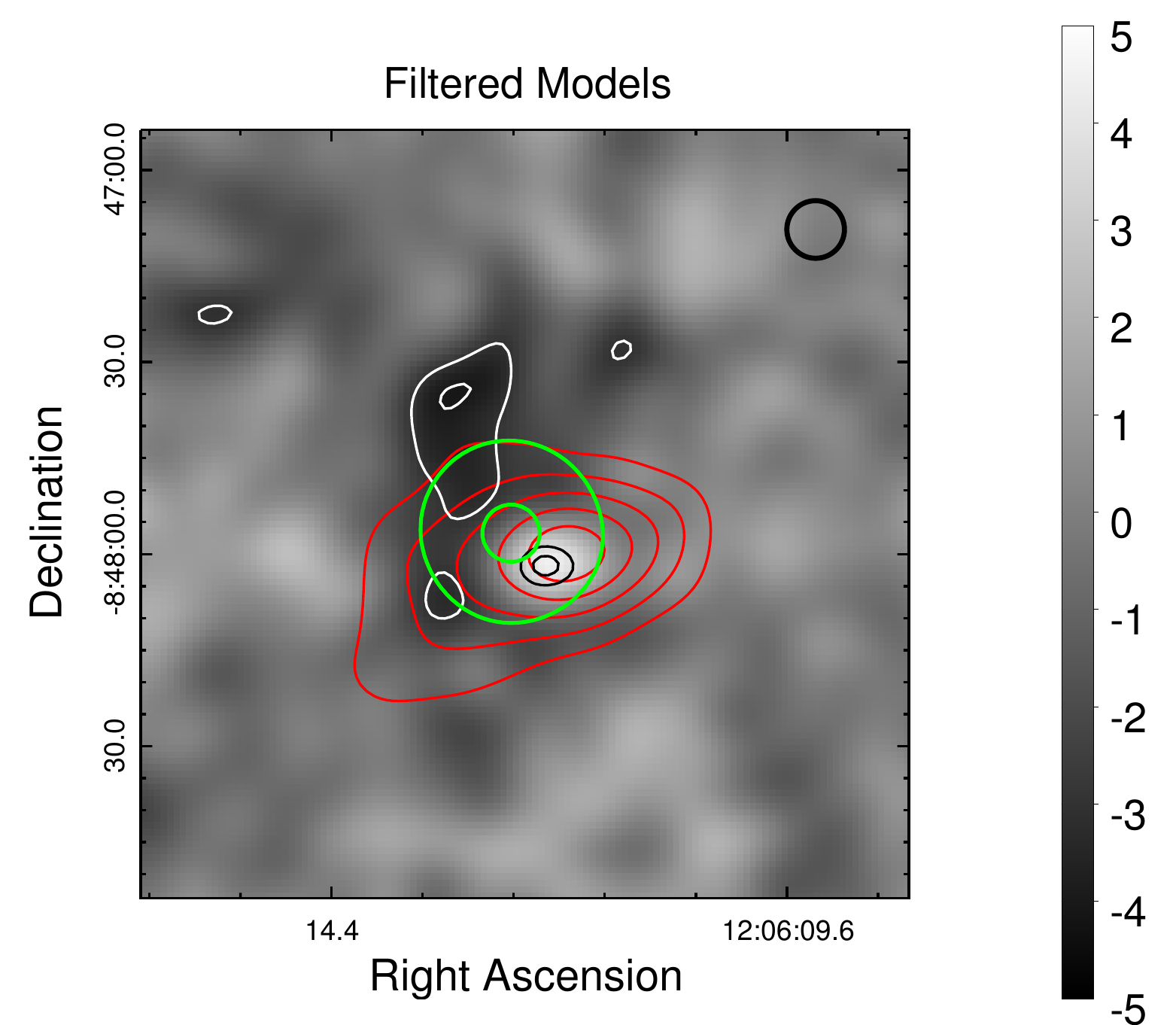}\\
    \includegraphics[width=3.1in]{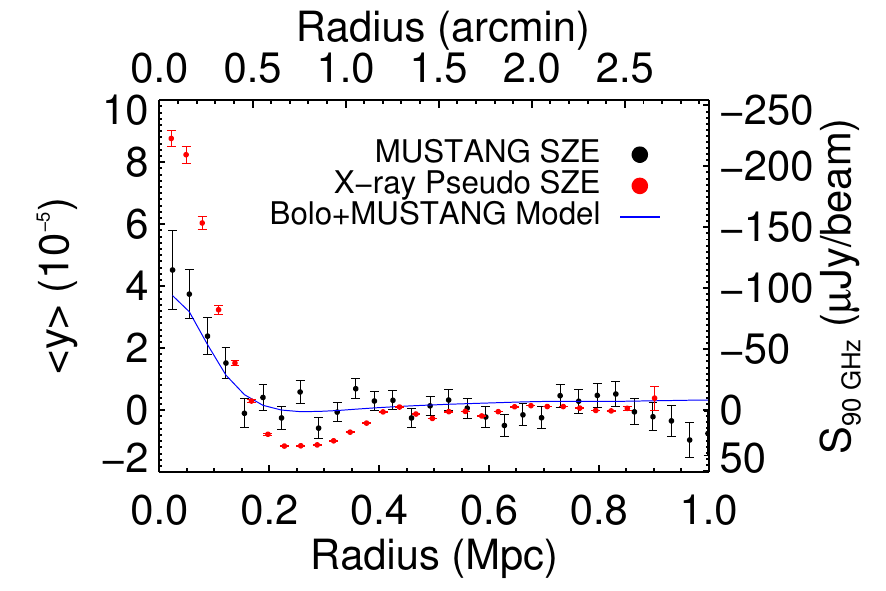}
  \caption{\footnotesize X-ray--derived (top, red contours) and
    Bolo+MUSTANG G7 model (top, green contours) for \macsb, each with
    the MUSTANG transfer function applied. Both have been smoothed
    with the MUSTANG PSF given by the black circle. The red and green
    contours start at $-75~\mu$Jy~beam$^{-1}$ and increase in steps of
    $-25~\mu$Jy~beam$^{-1}$. The MUSTANG S/N contours from
    Figure~\ref{fig:m1206snr} are overlaid in white. Azimuthally
    averaged radial profiles are shown in the lower panel.  The filtered
    X-ray derived flux shows a sharper peak relative to the Bolocam
    and MUSTANG data, which could be a result of the way in which the
    pseudo-$y$ map is normalized (see Figure~\ref{fig:m1206xray}).
    \label{fig:m1206inmod}}
\end{figure}

%% file: tab6.tex
\begin{deluxetable}{ccccc}
\tabletypesize{\scriptsize}
\tablecolumns{5}
\tablecaption{\macsb\ SZE Residual Flux and Lower Mass
Limits after Cluster Model and Point Source Subtraction \label{tbl:m1206fluxes}}
\tablehead{
Model(s)    & $S_{90}$     &  \Ysz\DA$^{2}$  & $M_{500}$\tablenotemark{a} & \Lx\tablenotemark{b}\\
Removed    &  ($\mu$Jy)  &  ($10^{-8}$\! Mpc$^{2}$)  & ($10^{13}$\! \msun) &($10^{43}$\! erg s$^{-1}$)}
\startdata
Point Src\tablenotemark{c} &  $-193\pm 36 $ & $32 \pm 6$   & $2.6 \pm 1.0$ & $3.0 \pm 1.7$ \\[.25pc]
G7+Pt\! Src                   & $-$\macsbflux  & $9.5 \pm 3.3$ & $1.3 \pm 0.7$  & $ 1.7 \pm 1.3$ 
\enddata
\tablecomments{Integrated flux estimates from the MUSTANG map. The
  first row corresponds to the total SZE flux with the point source
  emission taken into account. The bottom row is the residual flux
  after removing the best-fit G7 ICM model in addition to the point
  source flux. The integrated fluxes were computed within the regions
  enclosed by the 3-$\sigma$ contours shown in the right panel of
  Figure~\ref{fig:m1206snr}.}
\tablenotetext{a,b}{$M_{500}$ and \Lx\ are derived from the A10
  \Ysz-$M_{500}$ and $M_{500}$-\Lx\ scaling relations.}
\tablenotetext{c}{We use the $-765 \uJy$ point source model from
  Table~\ref{tbl:ptsrc}.}
\end{deluxetable}

%% file: fig13.tex
\begin{figure}[t!]
\begin{center}
    \includegraphics[width=3.25in]{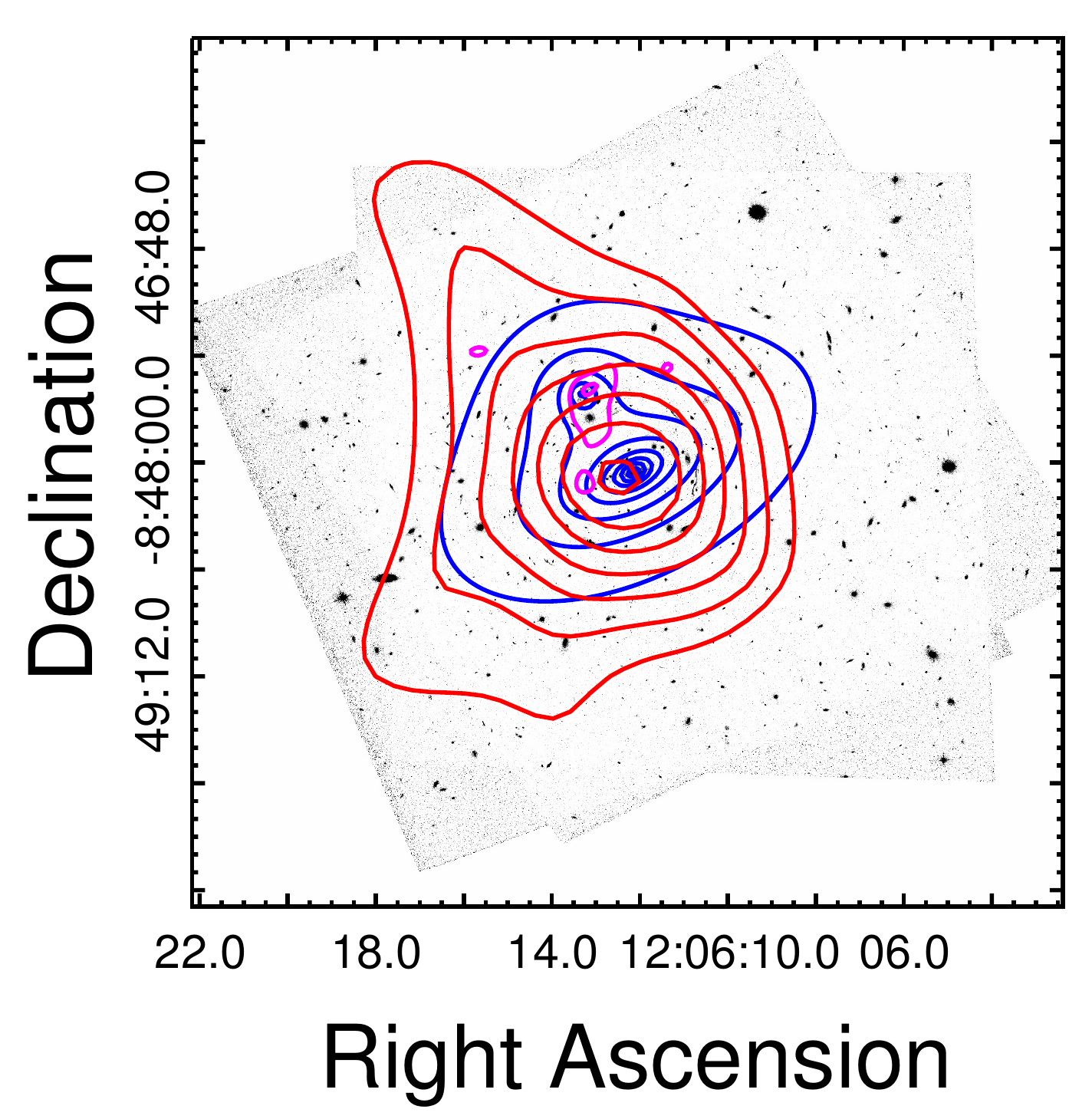}
\end{center}
  \caption{\footnotesize Optical image from HST (greyscale) overlaid
    with the weak-lensing mass distribution (red) from
    \citealt{umetsu2012}, the MUSTANG S/N contours (magenta), and the
    best fit eNFW+NFW two-halo model (blue). In addition to
    the E-W elongation noted in previous observations, there is an
    elongation to the NE. This suggests that the MUSTANG SZE detection
    may correlate with real structure such as an infalling galaxy
    group.
    \label{fig:m1206wl}}
\end{figure}

%% file: fig14.tex
\begin{figure}[t!]
\begin{center}
  \includegraphics[width=3.5in]{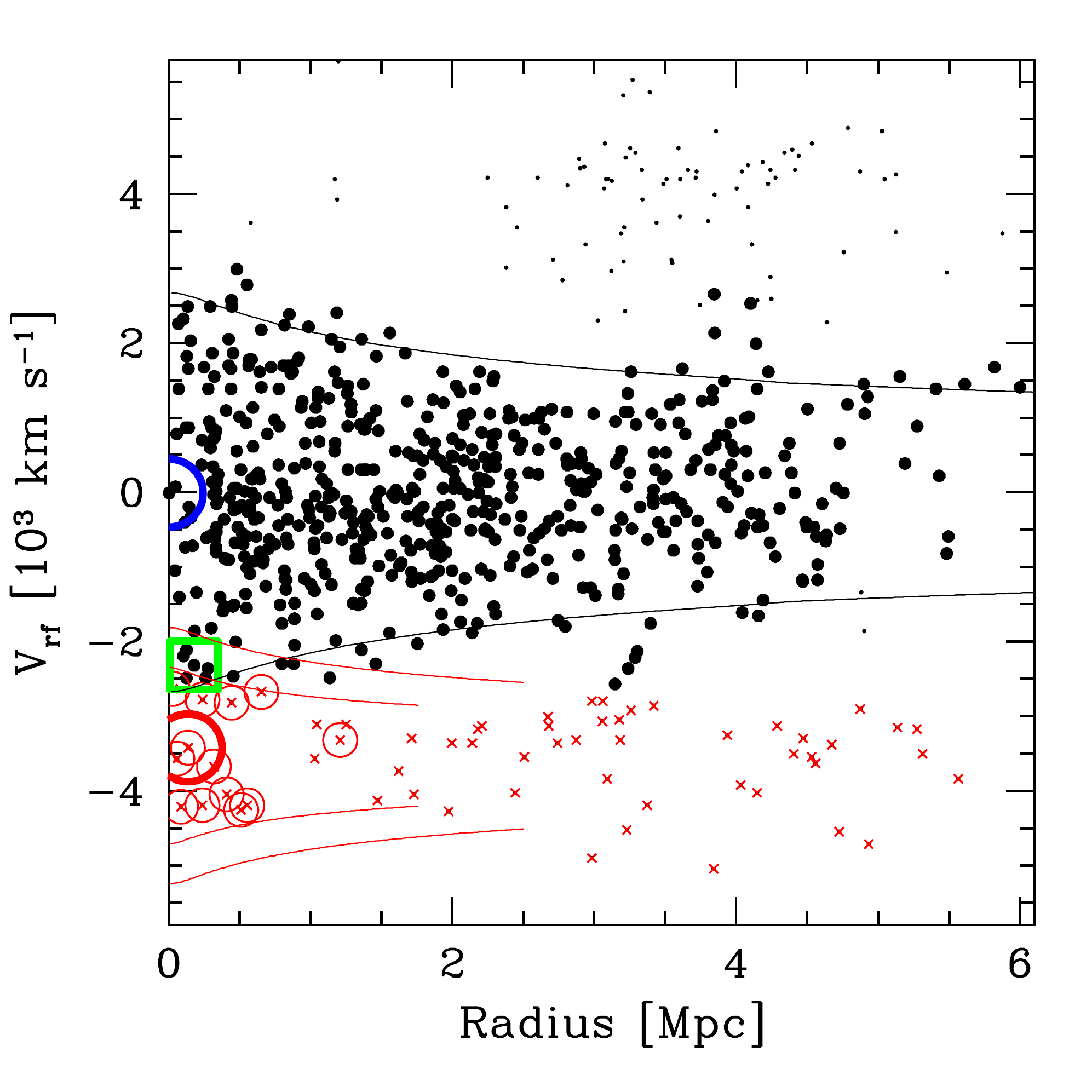}
\end{center}
  \caption{\footnotesize Rest-frame line-of-sight velocity versus
    projected distance from the center of \macsb, based on the
    CLASH-VLT VIMOS survey. The cluster center coincides with the
    position of its BCG (blue circle at 0,0). Solid black circles
    indicate cluster members (see also Figure~2 - lower panel of
    \citealt{biviano2013}).  Small red crosses indicate galaxies
    corresponding to the ${z\sim 0.42}$ peak in the redshift
    distribution and red circles highlight the 13 galaxies within
    $1.1$~Mpc of the SZE peak, that are the likely members of the
    putative group within $R_{200}$. The black and red curves show the
    limits due to the escape velocity in the cluster and the group,
    respectively. For the group we show both sets of curves
    corresponding to $M_{200}\pm 1\sigma$ error limits out to
    $2R_{200}$. The blue circle coincides with the cluster BCG, the
    red circle marks the brightest galaxy in the group which is
    located at the SZE peak, and the green square is a spiral galaxy
    that lies on the boundary between the cluster and the potential
    group.
    \label{fig:phase}}
\end{figure}

%% file: tab7.tex
\begin{deluxetable}{lc}
\tabletypesize{\scriptsize}
\tablewidth{0pt}
\tablecolumns{2}
\tablecaption{Group Mass Estimates}
\tablehead{
  Method	 & $M_{500}$         \\
       	         &  ($10^{14}$~\msun) 
} 
\startdata 
MUSTANG SZE             & $>$0.13  \\[.25pc]
X-ray                   & $<$4.5   \\[.25pc]
\sigV-$M$ (VLT) &  $1.4\pm0.9$  \\[.25pc]
eNFW+NFW                & $2.2\pm1.2$ \\
\enddata
\label{tbl:masses}
\tablecomments{Summary of the mass constraints and estimates derived
  from the SZE, X-ray, VLT, and weak-lensing data.}
\end{deluxetable}